\documentclass[twocolumn,showpacs,preprintnumbers,amsmath,amssymb]{revtex4-1}

\usepackage{CJK}

\usepackage{graphicx}

\usepackage{dcolumn}

\usepackage{bm}
\usepackage[normalem]{ulem}
\usepackage{color}
\usepackage{dsfont}

\usepackage{enumitem,kantlipsum}
\numberwithin{equation}{section}

\usepackage{epstopdf}

\newcommand{\bq}{\mathbf{q}}

\newcommand{\bv}{\mathbf{v}}

\newcommand{\br}{\mathbf{r}}
\newcommand{\bR}{\mathbf{R}}
\newcommand{\bS}{\mathbf{S}}

\newcommand{\bmu}{\boldsymbol{\mu}}
\newcommand{\bF}{\mathbf{F}}
\newcommand{\bff}{\mathbf{f}}
\newcommand{\bu}{\mathbf{u}}

\newcommand{\bw}{\mathbf{w}}

\newcommand{\hx}{\hat{x}}
\newcommand{\hy}{\hat{y}}
\newcommand{\hz}{\hat{z}}
\newcommand{\hn}{\hat{{\bf n}}}

\newcommand{\sep}{ \ \ \ , \ \ \ }

\newcommand{\beq}{\begin{equation}}
\newcommand{\eeq}{\end{equation}}
\newcommand{\beqn}{\begin{eqnarray}}
\newcommand{\eeqn}{\end{eqnarray}}
\newcommand{\pp}{\partial}
\newcommand{\dd}{{\rm d}}

\newcommand{\tdgl}{time-dependent Landau-Ginsburg-Wilson }
\newcommand{\kbt}{k_BT}

\newcommand{\cO}{{\cal O}}

\newcommand{\la}{\langle}

\newcommand{\ra}{\rangle}

\newcommand{\vnab}{{\bf \nabla}}

\newcommand{\bew}{\begin{widetext}}
\newcommand{\ew}{\end{widetext}}
\newcommand{\nn}{\nonumber}

\begin{document}

\title{Broken living layers: dislocations in active smectics}

\author{Frank Julicher}
\email{julicher@pks.mpg.de}
\affiliation{Max-Planck Institut f\"ur Physik Komplexer Systeme, N\"othnitzer 
Str. 38,
01187 Dresden, Germany}
\author{Jacques Prost}
\email{Jacques.Prost@curie.fr}
\affiliation{Mechanobiology Institute and Department of Biological Sciences, National University of Singapore, Singapore 117411}
\affiliation{3 Laboratoire Physico Chimie Curie, Institut Curie, PSL Research University, CNRS UMR168, 75005 Paris, France
}
\author{John Toner}
\email{jjt@uoregon.edu}
\affiliation{Department of Physics and Institute for Fundamental
	 Science, University of Oregon, Eugene, OR $97403$}

\begin{abstract}
We show that dislocations in active 2d smectics with underlying rotational symmetry are always unbound in the presence of noise, meaning the active smectic phase does not exist  for non-zero noise in $d=2$. The active smectic phase  can, like equilibrium smectics in 2d,  be stabilized by applying rotational symmetry breaking fields; however, even in the presence of such fields, active smectics are still much less stable against noise than equilibrium ones, when the symmetry breaking field(s) are weak.
\end{abstract} 
\maketitle

\section{Introduction\label{intro}}

Much of the richness of condensed matter physics is due to the great variety of possible different phases of matter. Each distinct phase breaks different symmetries of the underlying physical laws of the universe\cite{chaikin}.

One of the most interesting equilibrium phases of matter is the smectic A phase\cite{degennes}. This liquid crystalline phase is, as the term ``liquid crystal" suggests, a hybrid of a liquid and a crystalline solid. Specifically, a $d$-dimensional smectic A can be thought of as a one dimensional stack of $d-1$-dimensional isotropic fluids. In three dimensions,  this is a stack of two dimensional fluid layers. 

These fascinating phases exhibit a number of unique properties, including quasi-long-ranged order -i.e., algebraicly decaying translational correlations- in three dimensions\cite{caille}, and a breakdown of linearized hydrodynamics\cite{MRT}. 

{\it A priori}, all of the phases found in equilibrium could also be exhibited by active matter
\cite{Vicsek, TT1, TT2, TT3, TT4, TT5,Active4, Active1,Active2,Active3,JF1,JF2}
systems, in which the building blocks are  kept out of equilibrium by constant transduction of energy.  Examples of such systems are  living organisms, molecular motors, and robots,  to name just a few. In this paper, we consider active smectics A.

Whether or not a particular active phase is stable, and robust against noise, is, obviously, the first question one must ask about any potential phase of active matter.  Some phases of active matter - e.g., polar ordered active matter, of which a uniformly moving flock is the most obvious example, are actually more robust  than their equilibrium counterparts. Polar ordered flocks, for example, can exhibit {\it long}-ranged order in two dimensions, while their nearest equilibrium analog, ferromagnets, can exhibit only quasi-long ranged order  in two dimensions in the presence of noise - i.e., at finite temperature.

Other active phases, on the other hand, are less stable than their equilibrium counterparts. ``Wet" active nematics - that is, active nematics with momentum conservation -  are actually unstable\cite{simha}. 

So in this paper, we ask the question of whether or not active smectics are robust against noise. These systems have already been shown\cite{apolarsm, polarsm} to be stable at zero noise, and to be stable against noise if topological defects are ignored. These papers also asserted that these systems can be stable in two spatial dimensions in the presence of noise  against topological defects as well. Here, we show that this is not the case, 
 for the two simplest possible types of active smectic: dry Malthusian 
active smectics, and dry incompressible active smectics.  
We will now define these phases.

A dry Malthusian
active smectic is one in which nothing - not energy, not momentum, not even particle number\cite{TT5} - is conserved\cite{apolarsm, polarsm}. As a further simplification, we will consider only {\it apolar} smectics; that is, smectics with up-down symmetry (that is, symmetry between the two directions normal to the smectic layers).

 A  dry incompressible active smectic is simply an active smectic  with no momentum conservation whose {\it mean} density is fixed. By ``mean" density, we mean the small wavelength components of the density; the components at $q_0\equiv {2n\pi\over a}$, with $a$ the smectic layer spacing,  are non-zero for all integer $n$ as a result of the spontaneous density modulation that defines the smectic.

Such completely non-conservative systems are by no means of purely academic interest. Tissues growing or developing on a substrate \cite{tissue1,tissue}, for example, lack all conservation laws: cell number is not conserved due to cell division and death. Momentum is lost due to friction with the substrate, and can be gained due to active forces between the cells and that substrate. Finally, energy is not conserved, both due to friction with the substrate, and because living cells have a fuel source (i.e., food) which enables them to 
consume energy. Hence, if such a system was thin compared to its lateral extent, and spontaneously
formed a layered structure, it would be a two dimensional dry Malthusian
active smectic of exactly the type we describe here.   However, smectic order has not been observed experimentally yet in two-dimensional tissues.

Interestingly, the cell actomyosin system may provide an example of  a Malthusian active smectic. Cell contractility is known to result mainly from the action of the myosin II molecular motors on the cortical actin gel. Myosin II motors have a long tail, and two active heads. The tails of several motors tend to bundle in a way similar to the tail bundling of phospholipids, but at a different scale: the resulting head to head distance is on the order of 300 nm\cite{myosins}. Often the picture of the bundle is that of a symmetric flower bouquet and their distribution in the actin gel is essentially random. More ordered structures exist: there is well defined 3d crystalline order in muscles, and 1d periodic patterns occur in stress fibers. An intermediate arrangement is observed in lamellipodia of fibroblasts \cite{myosins,myosins1,myosins2}: the myosin tails arrange in lines separated from the heads, building a clear 2d smectic order. The actin filaments, above this structure,  are on average orthogonal to the myosin lines with no specific translational order. The  bundling/unbundling process does not conserve myosin number and interaction with the substrate  exchange momentum. Hence, this is a "dry" system in the sense defined earlier. Such systems of stacked myosin II filaments in cells have thus the characteristics of dry active Malthusian smectics.

In addition, we think our results shed considerable light on the dislocation behavior one might expect in more complex active smectic systems in which one or more quantities are conserved. For instance, our results are valid in 2d incompressible systems with boundary conditions allowing for layer number variation as shown in the last section of this manuscript.
This is for instance the case for the roll structures in Rayleigh-Benard instabilities\cite{1st}.

We find that, despite being more stable in spin-wave theory - that is, when topological defects (i.e., dislocations), are neglected,  active smectics in rotation invariant environments (which we will hereafter refer to by the shorthand ``rotation invariant active smectics") are unstable against dislocation unbinding in the presence of {\it any} noise, no matter how small. Furthermore, although they can be stabilized by rotational symmetry breaking fields, they are still less stable than equilibrium smectics in symmetry breaking fields of the same strength. Specifically, in the active smectics we study here, we'll define $\Delta_c$ as the critical value of the noise strength $\Delta$   above which dislocations unbind, causing the smectic to melt into an active nematic. This critical value $\Delta_c$ grows {\it linearly} with the applied symmetry breaking field strength $g$ for small $g$; that is
\beq
\Delta_c\propto g \,\,\, {\rm as} \,\,\,\,  g\to0\,.
\label{noisescaleintro}
\eeq
This result should be contrasted with the equilibrium result\cite{KT} for the transition temperature $T_c$, (temperature is the equilibrium analog of the noise strength in active smectics):
\beq
T_c^{\rm eq}\propto \sqrt{g} \,\,\, {\rm as} \,\,\,\,  g\to0\,,
\label{noisescale eq}
\eeq
whose derivation we'll review in section \ref{symbreakeq}. We therefore see that, for small symmetry breaking fields $g$, the critical noise strength $\Delta_c$ for dislocation unbinding and the melting of the smectic phase is much smaller for active smectics than for equilibrium smectics. That is, active smectics are {\it less} robust against melting, even in the presence of symmetry breaking fields, than their equilibrium counterparts.

Like equilibrium smectics in the presence of a rotational symmetry breaking field, active smectics in such a symmetry breaking field exhibit quasi-long-ranged translational correlations for noise strength smaller than the critical value. This is most clearly manifest in the Fourier transformed density-density correlation function (which is also the X-ray structure factor in scattering experiments). This exhibits peaks at wavevectors $\bq_n=nq_0\hy+{\bf \delta q}$, where $q_0={2\pi\over a}$, with $a$ the smectic layer spacing. Near these peaks, we have 
\beq
\langle |\rho(\bq,t)|^2\rangle\propto |{\bf \delta q}|^{-2+n^2\eta(g, \Delta)} \,,
\label{Braggeqsmec}
\eeq
with $\eta(g, \Delta)$ a non-universal exponent that depends on the symmetry breaking field strength $g$ and the noise strength $\Delta$ (as well as other smectic parameters). 

Another consequence of our result   (\ref{noisescaleintro}) is that the critical value $\eta_c$ of the exponent $\eta$  vanishes linearly with symmetry breaking field:
\beq
\eta_c\propto g \,\,\, {\rm as} \,\,\,\,  g\to0\,.
\label{etascaleintro}
\eeq
in contrast to equilibrium smectics, for which $\eta_c=1/4$, universally, independent of the applied symmetry breaking field.

The remainder of this paper is organized as follows: in section \ref{eq}, we review the theory of equilibrium smectics, both in rotation-invariant systems \ref{rotinveq} and with rotational symmetry breaking fields \ref{symbreakeq}. In section \ref{swact}, we review the spin-wave theory of active smectics (that is, the theory in which dislocations are neglected). Section
\ref{disconf} presents the calculation of the fields due to dislocations, which prove to be identical in form to those found in equilibrium smectics in the presence of a rotational symmetry breaking field. Section \ref{diseom}
derives the equation of motion for dislocations in an active, rotation invariant smectic. In section \ref{homeo}, we show that this equation of motion leads, for fixed boundary conditions, to the achievement of a type of ``homeostasis,", in which isolated dislocations do not spontaneously
move. For ``constant stress" boundary conditions, on the other hand, the smectic will either collapse, or grow without bound. In section
\ref{dis unbind}, we show that even in the homeostatic state, in the presence of noise, dislocations are always unbound in rotation invariant smectics.
Rotational symmetry breaking fields can 
stabilize smectic quasi-long-ranged order in active smectics, as we show in
\ref{dis sym break}. We also show in that section that although symmetry breaking fields do stabilize smectic order, the resultant order is still very weak for small fields, in the
sense that it can be much more easily destroyed by noise than in equilibrium smectics.
We then generalize these results to the case of incompressible dry active smectics.
Finally, in section
\ref{sum}, we summarize our results, speculate about the behavior of more complicated smectic systems with conservation laws, and suggest avenues for further work.

\section{Review of Equilibrium 2d smectics}\label{eq}
  \subsection{Rotation invariant 2d equilibrium smectics}\label{rotinveq}
   
\subsubsection{"Spin wave" (Phonon) theory}\label{sweq}
\begin{figure}
\begin{center}
\includegraphics[scale=.35]{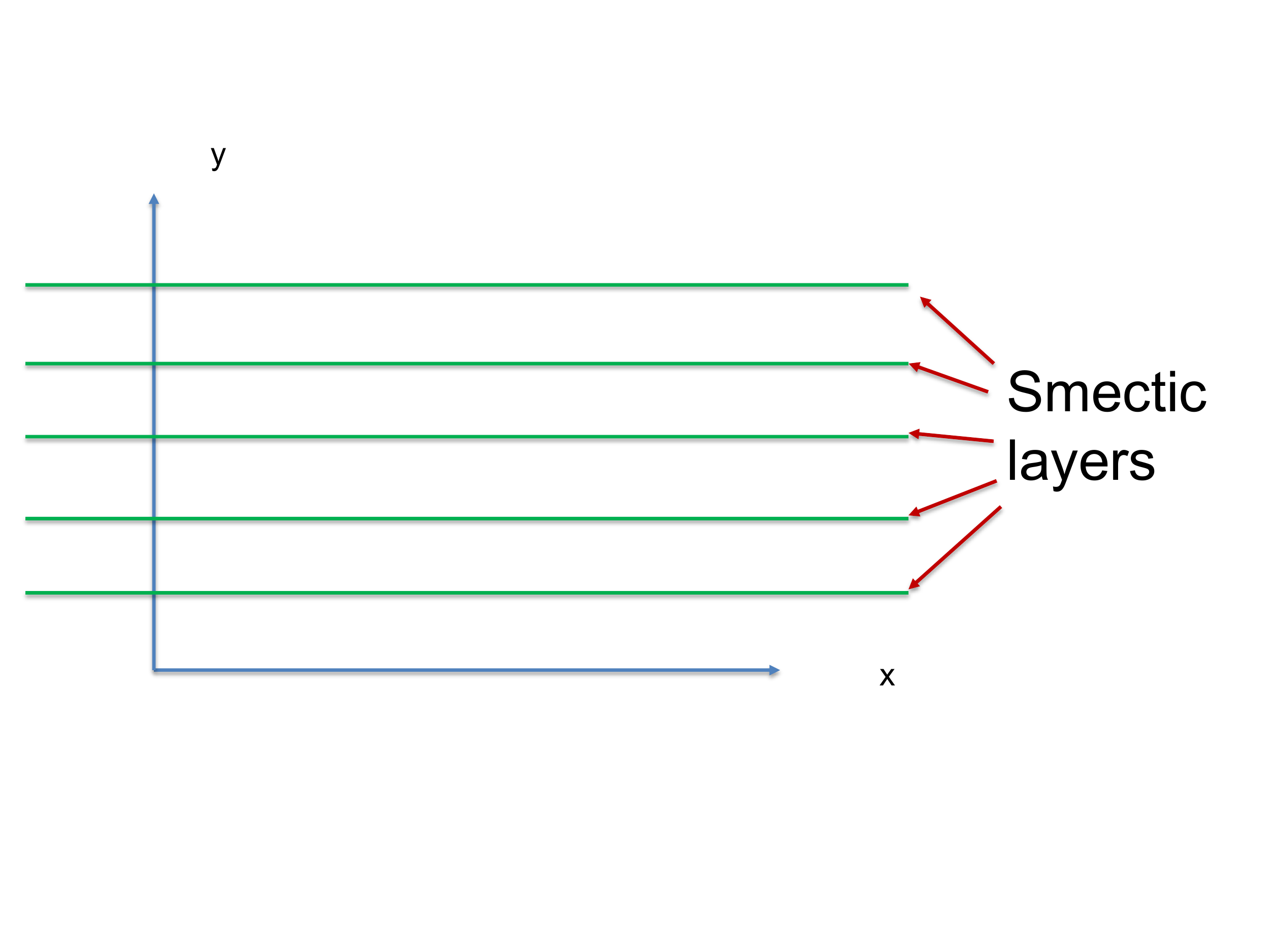}
\end{center}
\caption{Schematic of the ideal smectic state, in which the layers are parallel, and uniformly spaced. We choose our coordinates $(x,y)$ so that the $x$-axis runs parallel to the layers, as shown. }
\label{smectic}
\end{figure}

Any smectic A phase (either equilibrium or active)  is characterized by the spontaneous breaking of translational invariance in one direction (in contrast to a crystal, in which translational invariance is broken in all $d$ dimensions, where $d$ is the dimension of space). This is equivalent to saying that the system spontaneously layers, with the additional requirement that the layers are "liquid-like", in the sense of being homogeneous along the layers. We will choose our co-ordinates throughout this paper so that the direction in which translational invariance is broken is $y$. This means the layers, in the absence of fluctuations, run parallel to $x$ (see  figure [\ref{smectic}]). 

\begin{figure}
\begin{center}
\includegraphics[scale=.35]{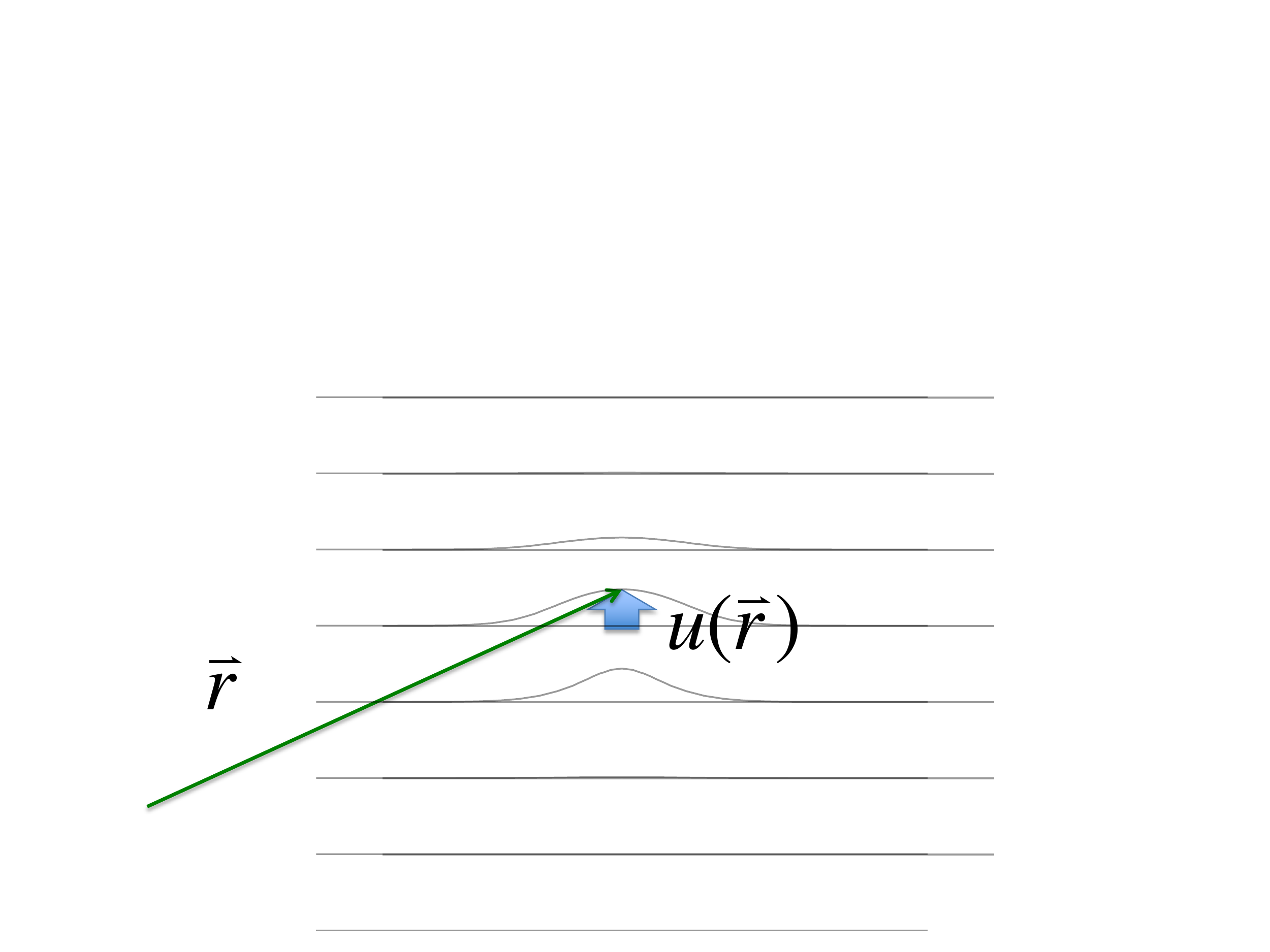}
\end{center}
\caption{Definition of the layer displacement field $u(\br,t)$. The straight parallel line are the reference positions of the layers, while the curved lines depict a fluctuation in the layer positions. The layer displacement field $u(\br,t)$ is the distance from the reference position to the fluctuating position of the layers in the vicinity of the spatial point $\br$, as illustrated.}
\label{udeffig}
\end{figure}

Since the smectic breaks the continuous translational symmetry of free space, such systems (again whether equilibrium or active) have a "Goldstone mode" associated with the breaking of this symmetry. In smectics, we usually take this Goldstone mode to be the local displacement $u(\br,t)$ of the layers in the vicinity of the spatial point $\br$ away from some reference set of parallel, regularly spaced layer positions (see figure [\ref{udeffig}]).

To describe such systems in equilibrium, one introduces\cite{caille, chaikin, degennes} a phenomenological elastic Hamiltonian (sometimes called the elastic free energy). This is constructed as an expansion in powers of spatial gradients of the
  displacement field $u(\br,t)$, keeping all terms to leading order in spatial gradients allowed by the symmetries of the system. Translational invariance requires that all terms involve at least one spatial derivative of $u(\br,t)$, since a spatially uniform displacement (i.e., $u(\br,t)={\rm constant}$) must cost no energy.
 
  {\it Rotation} invariance is somewhat more subtle. Its implications can be understood by recognizing that a uniform rotation of the layers by a small angle $\phi\ll1$ can be represented in our coordinate system by a {\it non-uniform} displacement field
\beq
u(\br, t)=\phi x \,.
\label{rotation}
\eeq  

From this expression, we see that 
\beq
\pp_x u=\phi \,,
\label{rotation dif}
\eeq
that is, the $x$-derivative of the displacement field $u$ gives the rotation angle of the layers away from their reference orientation.

The relation \eqref{rotation dif} continues to apply for arbitrary layer distortions; that is, the $x$-derivative of $u$ locally gives the local tilt of the layers away from their reference orientation (provided that tilt is small, of course). We will make much use of this relationship throughout this paper.

Rotation invariance therefore {\it forbids} the inclusion of terms that depend on $\pp_x u$ in the elastic Hamiltonian, since such terms will be non-zero for the uniform rotation \eqref{rotation}. Therefore, the leading order term involving $x$-derivatives of $u(\br,t)$ in $H$ is a term proportional to $(\pp_x^2 u)^2$, which represents the energy cost of {\it bending } the layers.

There is no such prohibition against terms involving $\pp_yu$. Indeed, a term proportional to $(\pp_yu)^2$ can easily be seen to represent the energy cost of {\it compressing} the layers closer together (for  $\pp_y u<0$) or stretching them further apart (for  $\pp_y u>0$). It is straightforward to show that
\beq
\delta a=a\pp_yu  \,,
\label{del a}
\eeq
where $\delta a$ is the departure of the local layer spacing from its energetically optimal value $a$. This is another relation which we will use repeatedly throughout this paper.

These considerations lead, to quadratic order in $u$, to the elastic Hamiltonian\cite{ caille, chaikin, degennes}:
  \beq
H_{\rm sm}=\frac{1}{2} \int \dd^2r \left[B(\pp_y u)^2+K(\pp_x^2 u)^2\right]\, .\label{Hsm rot inv}
\eeq
While higher than quadratic order in $u$ terms are actually important in 2d equilibrium smectics \cite{Golub}, we will not consider them here, since they do not play any role in active apolar smectics\cite{polarsm, polar smectic}.

The simplest, purely relaxational, equilibrium dynamics associated with this Hamiltonian is the \tdgl  equation of motion:
\beq
 \pp_t  u=-\Gamma{\delta H_{\rm sm}\over\delta u}+f_{\rm u} \,.
 \label{TDGL}
 \eeq
where $f_{\rm u}(\br,t)$ is a Gaussian, zero mean white noise that drives the smectic to thermal equilibrium, governed by the Hamiltonian $H_{\rm sm}$ at temperature $T$. To do this, its variance must obey the ``fluctuation-dissipation theorem"\cite{chaikin}, which requires
\begin{equation}
 \langle f_u ({\bf r},t)f_{u}({\bf r}',t')\rangle=2 \Gamma k_BT\delta(
 {\bf r}-{\bf r}') 
\delta(t-t')\,.
\label{sw noise sm1}
\end{equation}

Using the Hamiltonian \eqref{Hsm rot inv} in the equation of motion \eqref{TDGL}  gives:  \begin{eqnarray}
\partial_t u = \Gamma B \partial_z^2 u  -
\Gamma K\pp_x^4 u + f_{u} \,.
\label{u eq rot inv EOM}
\end{eqnarray}

Note that this equation exhibits {\it subdiffusive} behavior in the $x$-direction. That is, relaxation along the $x$-direction is even slower than diffusive; specifically, the lifetime of a plane wave $u$ field running along $x$ with wavelength $\lambda$ grows like $\lambda^4$ for large $\lambda$, in contrast to the $\lambda^2$ scaling of simple diffusion.

This slowness of response to distortions in the $x$-direction, like the corresponding "softness" in the $x$-direction of the elastic Hamiltonian \eqref{Hsm rot inv}, is a direct consequence of the rotation invariance (i.e., the zero energy cost of pure rotations \eqref{rotation}) discussed earlier.

This rotation invariance can be removed by applying an external symmetry breaking field, as we'll discuss in subsection B.

   \subsubsection{Dislocation effects: There are no 2d equilibrium smectics at $T\ne0$}\label{disloeq}

The preceding discussion constitutes what is normally called "spin-wave" theory. It assumes that it is possible to define throughout the entire system a unique, single valued displacement field $u(\br,t)$ throughout the system. This is in fact not the case if dislocations (see figure [\ref{Burgers}]) are present in the system. 

We will now review the theory of these defects in equilibrium, as first developed by Pershan\cite{pershan}. 

\begin{figure}
\begin{center}
\includegraphics[scale=.35]{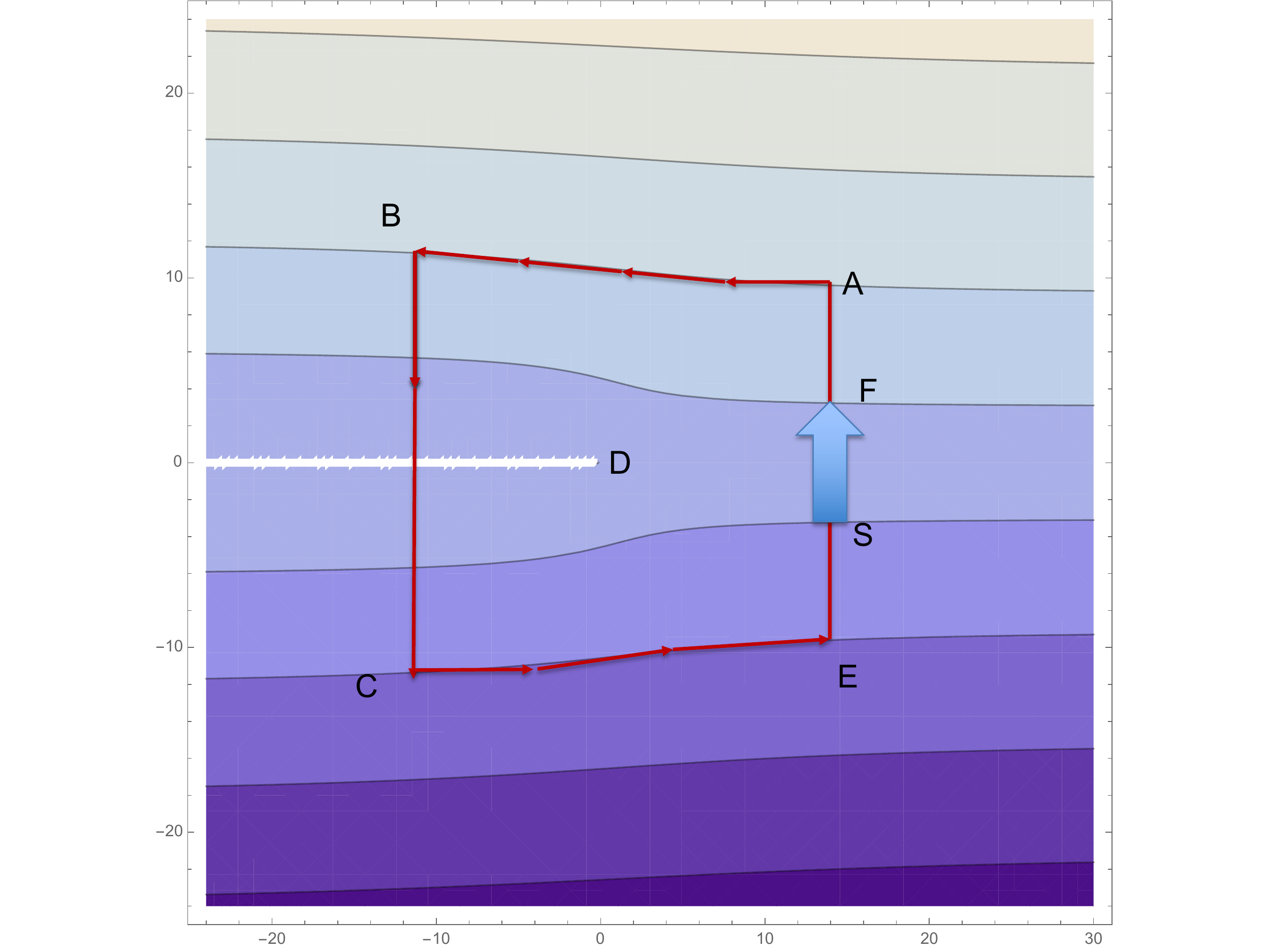}
\end{center}
\caption{Illustration of the Burgers' construction for dislocations described in the text.}
\label{Burgers'}
\end{figure}

As can easily be seen from figure [\ref{Burgers'}], one can think of a dislocation in a smectic as a place where a layer ends. That these "topological defects" make it impossible to define a single valued $u$ field can be seen by the well-known "Burgers' construction", which is also illustrated in figure \ref{Burgers'}. In this construction,  one  "walks" in a closed loop around the dislocation, crossing equal numbers of layers while moving up and moving down. In the example of figure \ref{Burgers'}, one crosses two layers while going up from the starting point S to the corner A, plus another two while moving from the lower right corner E to the final point F, for a total of four going up, and crosses four going down between the upper left corner B and the lower left corner C. Clearly, if the dislocation (at point D in figure \ref{Burgers'}) were not there,
or if the path did not enclose the dislocation (i.e., the point D), this path would have returned to the starting point S. Equally clearly, when it {\it does} enclose the dislocation, the path does {\it not} return to $S$, but, rather, overshoots closure by the length of the thick blue arrow between S and F, whose length is clearly the layer spacing $a$. This failure to close is known as the "Burgers' number" $b$ for the dislocation, and in this case is $b=+1$.  

It is straightforward to see that in general, the Burgers' number will be an integer, the integer being simply the difference between the number of layers coming in from the left that end inside the loop, and the number coming in from the right that do so.

The Burgers' number, defined in this way, is the analog for smectics (which, we remind the reader, only translationally order in {\it one} direction) to the better-known "Burgers' vector" defined by an almost identical construction in crystalline solids (which translationally order in {\it all} directions).

One way of thinking about this result is that, if we had defined the displacement field at the starting point S to be $u(S)=0$, we would, after having 
completed the loop, 
found that the layer had been displaced up by an amount $na$ for $b=n$. That is, $u$ is no longer single valued, but instead increases by $ba$ every time one moves around a loop enclosing the dislocation.
Mathematically, the contour integral
$\int_C m \hat {\bf n} \cdot {\bf dl}$ counts the number of layers traversed along the integration path $C$,
where $m=1/a (\br)$ is the local layer density   at the point $\br$, and $\hat n (\br)$ is the unit normal to the layers   at the point $\br$.
The statement that $u$ is not single valued, but changes when moving around a loop enclosing dislocations,   is equivalent to the statement that a nonvanishing number of layers   is encountered    by a closed loop
\beq
\oint m \hat {\bf n} \cdot {\bf dl}=\sum_\alpha b_\alpha \,.
\label{Burgers int}
\eeq
Here  we have now generalized to the case of many dislocations, labeled by $\alpha$, with Burgers' numbers $b_\alpha$, at positions $\br_\alpha$ that are enclosed by the loop over which the contour integral on the left-hand side of \eqref{Burgers int} is done. We reiterate that the Burgers' number $b_\alpha$ of each dislocation must be an integer that is, $b_\alpha=n$, with $n$ an integer.

Applying Stokes' theorem to \eqref{Burgers int} gives
\beq
\nabla\times m \hat {\bf n}=\sum_\alpha b_\alpha \delta(\br-\br_\alpha) \,,
\label{burgers dif}
\eeq
In \eqref{burgers dif}, we have defined the curl in two dimensions in the usual way; that is, as a scalar given, for any vector $\bv$,  by $\nabla\times\bv\equiv\pp_xv_y-\pp_yv_x$.

It is convenient to define  ${\bf w}=(a_0 m-1) \hat {\bf n}$, where $a_0$ is the layer spacing in a reference state. For small displacements it can be written as
\beq
\bw(\br,t)\simeq\phi(\br,t)\hx+\left({\delta a(\br,t)\over a_0}\right) \hy \,.
\label{wdef}
\eeq
with $\phi(\br,t)$ and $\delta a(\br,t)$ respectively the local tilt of the layers at
$\br$ at time $t$, and the local change in the layer spacing.

Keeping in mind our earlier discussion of the relationships (\ref{rotation dif}) and (\ref{del a}) which hold between these two quantities $\phi(\br,t)$ and $\delta a(\br,t)$ and the layer displacement field $u(\br,t)$ in the absence of dislocations, we see that $\bw$ is clearly simply the generalization of $\nabla u$ to situations in which dislocations are present. To say this another way, when dislocations are {\it absent},  
\beq
\bw=\nabla u \sep {\rm no} \,{\rm dislocations}\,.
\label{w no dis}
\eeq
Thus, $\bw$ is the natural generalization of the vector field $\nabla u$ in the presence of dislocations, see Appendix \ref{appendix}. 

We can use this idea to calculate the $\bw$ field of a dislocation. That will be the field that minimizes the energy of the system for a given configuration of dislocations. That is, we wish to find the field $\bw(\br)$ that minimizes the energy of the system subject to the constraint 
\beq
\nabla\times\bw(\br)=\sum_\alpha a_0 b_\alpha \delta(\br-\br_\alpha) \,,
\label{burgers dif w}
\eeq
which is where the dislocation configuration $\{b_\alpha, \br_\alpha\}$ enters the calculation.

In the absence of dislocations, the minimum energy configuration $u(\br)$ can be obtained from the Euler-Lagrange equation associated with the smectic elastic Hamiltonian (\ref{Hsm rot inv}). That equation is easily seen to be:
\beq
B\pp^2_yu(\br)-K\partial^4_xu(\br)=0 \,.
\label{ELG rot inv}
\eeq
We can obviously rewrite this as
\beq
B\pp_y(\pp_yu(\br))-K\partial^3_x(\pp_xu(\br))=0 \,.
\label{ELG rot inv2}
\eeq
Now generalizing this equation to situations in which dislocations are present by replacing $\nabla u$ with $\bw$ gives
\beq
B\pp_yw_y(\br)-K\partial^3_xw_x(\br)=0 \,.
\label{ELG w rot inv}
\eeq
 
The other condition on dislocations is the Burgers' condition \eqref{burgers dif w}.  

These two simultaneous linear  equations \eqref{ELG w rot inv} and \eqref{burgers dif w} 
can  be easily solved by
Fourier transforming in space;   
this gives
\beq
Bq_yw_y(\bq)+Kq_x^3w_x(\bq)=0 \,,
\label{steady state FT}
\eeq
and
\beq
q_xw_y(\bq)-q_yw_x(\bq)=-i\sum_\alpha a_0b_\alpha e^{-i\bq\cdot\br_\alpha}\,,
\label{BurgerFT}
\eeq
Solving these simple linear equations gives 
\beqn
&&w_x(\bq)={iBq_y \over Bq_y^2+Kq_x^4}\sum_\alpha a_0 b_\alpha e^{-i\bq\cdot\br_\alpha} \,,\\
&&w_y(\bq)={-iKq_x^3\over Bq_y^2+Kq_x^4}\sum_\alpha a_0 b_\alpha e^{-i\bq\cdot\br_\alpha} \,.
\label{w solFT rot inv}
\eeqn

Fourier transforming these solutions back to real space gives
\beqn
&&w_x(\br)=\sum_{\alpha} a_0 b_\alpha  G_x(\br-\br_\alpha) \,,\\
&&w_y(\br)=\sum_{\alpha} a_0 b_\alpha  G_y(\br-\br_\alpha) \,,
\label{w rot inv}
\eeqn
where the Green's functions $G_{x,y}$ are given by
\beqn
&&G_x(\br)=-{1\over4\sqrt{\pi\lambda |y|}}\exp\bigg[-\left({x^2\over4\lambda |y|}\right)\bigg]  {\rm sgn}(y)\,,\nonumber\\\\
&&G_y(\br)={x\over8\sqrt{\pi\lambda |y|^3}}\exp\bigg[-\left({x^2\over4\lambda |y|}\right)\bigg] \,,\nonumber\\
\label{G disloc iso smec}
\eeqn
where we've defined $\lambda\equiv(K/B)^{1/2}$.

We can
also obtain the energy of interaction between dislocations by inserting the solution \eqref{w solFT rot inv} for $\bw$ into 
our elastic Hamiltonian \eqref{Hsm rot inv}. To do so, we must first rewrite \eqref{Hsm rot inv}  in terms of $\bw$ using the same replacement $\nabla u\to\bw$ we've been using. This gives:
\beq
H_{\rm sm}=\frac{1}{2} \int \dd^2r \left[Bw_y^2+K(\pp_x w_x)^2\right]\, .\label{Hsmw rot inv w}
\eeq
Fourier transforming this,  inserting the solution \eqref{w solFT rot inv} for $\bw$ into the result, and Fourier transforming back to real space gives
\beq
H(\{b_\alpha, \br_\alpha\})=\sum_{\alpha,\beta} a_0^2 b_\alpha b_\beta U(\br_\alpha-\br_\beta)
\label{disloc H}
\eeq
where the pairwise interaction potential 
\beq
U(\br)={B\over4}\sqrt{\lambda\over\pi |y|}\exp\bigg[-\left({x^2\over4\lambda |y|}\right)\bigg] \,.
\label{U disloc iso smec}
\eeq

Because this vanishes as the separation between dislocations goes to infinity, dislocations will always be unbound in smectics at any non-zero temperature, thereby destroying the smectic order. This means that 2d smectics do not actually exist in rotation invariant systems at non-zero temperature, as first shown (using exactly this argument)  by \cite{1st}.

While we do not need the equation of motion for the dislocations in this equilibrium system, since the statistical mechanics is determined entirely by the Hamiltonian \eqref{Hsm rot inv}, it is instructive to formulate those equations of motion. This will allow us later to compare and contrast them with the equations of motion for dislocations in active smectics, for which the equation of motion is the only information we have about the behavior of dislocations in those non-equilibrium systems.

To obtain the equations of motion, we first calculate the forces arising from the potential \eqref{U disloc iso smec}.

Consider a system with only two dislocations $\alpha=(1,2)$ of Burgers' charges:
$b_1$ and $b_2$, at positions $\br_1={\bf 0}$, $\br_2=\br=x\hx+y\hy$.
The dislocation Hamiltonian \eqref{disloc H} then implies that the energy of this pair will be
\beq
V(\br)={Bb_1b_2a_0^2\over4}\sqrt{\lambda\over\pi |y|}\exp\bigg[-\left({x^2\over4\lambda |y|}\right)\bigg] \,.
\label{U disloc iso smec pair}
\eeq
The force experienced by dislocation $\alpha=2$
    will therefore be  ${\bf F}=F_x\hx+F_y\hy$, with its Cartesian components $F_{x,y}$ given by
\bew
 \begin{eqnarray}
F_x&=&-\pp_xV(\br)={Bb_1b_2a_0^2x\over8\sqrt{\pi\lambda |y|^3}}\exp\bigg[-\left({x^2\over4\lambda |y|}\right)\bigg]=Bb_2a_0w_y^{(1)}(\br)
\nonumber\\
F_y&=&-\pp_yV(\br)=-{Bb_1b_2a_0^2{\rm sgn}(y)\over16\sqrt{\pi\lambda |y|^5}}\bigg\{x^2-2\lambda|y|\bigg\}\exp\bigg[-\left({x^2\over4\lambda |y|}\right)\bigg]=Kb_2a_0\pp_x^2w_x^{(1)}(\br)
\label{rot inv eq forces}
 \end{eqnarray}
\ew
where by $\bw^{1}(\br)$, we mean the contribution to the field $\bw$ at the position $\br$ of the $\alpha=2$ dislocation coming from the $\alpha=1$ dislocation (i.e., neglecting the the field created by dislocation $\alpha=2$ itself). It is straightforward to show that the generalization of these forces to configurations with more than two dislocations is:
 \begin{eqnarray}
F_x^\alpha&=&Bb_\alpha a_0w_y(\br_\alpha)
\nonumber\\
F_y^\alpha&=&Kb_\alpha a_0\pp_x^2w_x(\br_\alpha)
\label{rot inv eq forces gen}
 \end{eqnarray}
where by $\bw(\br_\alpha)$, we mean the contribution to the field $\bw$ at the position $\br_\alpha$ of the $\alpha$'th dislocation coming from all of the other dislocations (i.e., neglecting the  field created by dislocation $\alpha$ itself).

Since the dislocation cannot "tell" whether the local field $\bw$ is created by other dislocations, by spin waves, or by externally applied stresses, we expect \eqref{rot inv eq forces gen} to hold more generally if we take $\bw_\alpha$ on the right-hand side of those equations to be the entire $\bw$ field, {\it excluding} the part due to dislocation $\alpha$ itself.
 This proves to be important when we consider the effect of stresses at the boundary on dislocation   motion.

There are two important features of the result \eqref{rot inv eq forces gen} that should be noted:

\noindent 1) the force on a given dislocation $\alpha$ is determined entirely by the {\it local} value of the field $\bw(\br_\alpha)$ (and its derivatives) at the location $\br_\alpha$ of that dislocation  (excluding the part of that field due to the given dislocation itself).

\noindent 2) The dependence of the force on the $x$ component $w_x$ involves spatial derivatives of that component; a uniform $w_x$ does not generate any force on the dislocation. This is a consequence of rotation invariance: as shown by equation \eqref{rot inv eq forces gen}, a spatially uniform $w_x$ corresponds to a uniform rotation of the layers, which clearly {\it cannot} lead to any force on the dislocation in a rotation-invariant system. This consideration will continue to apply in active smectics, and will forbid certain terms in the force in those systems which one might otherwise expect, as we will see in section [\ref{diseom}] below.

Since there will be friction between the dislocations and the underlying substrate, we expect the {\it velocity}  $\bv$ (rather than the {\it acceleration} $\dot{\bv}$)  of the dislocations to be linear in the 
force $\bF$. That is,
\beq
\bv_\alpha=\bmu\bF_\alpha \,,
\label{mobtens}
\eeq

where $\bmu$ is a constant "mobility tensor". On symmetry grounds, we expect this tensor to be diagonal in our $(x,y)$ coordinate system (i.e., with the $x$ and $y$ axes respectively parallel
and perpendicular to the mean smectic layers); hence
\beq
v_x^\alpha=\mu_x F_x^\alpha \,, \, v_y^\alpha=\mu_y F_y^\alpha \,.
\label{vdisiso}
\eeq

Using our earlier results \eqref{rot inv eq forces gen} for the forces on the dislocations, we can rewrite these as
 \begin{eqnarray}
v_x^\alpha&=&\mu_xBb_\alpha a_0 w_y(\br_\alpha)
\nonumber\\
v_y^\alpha&=&Kb_\alpha a_0\pp_x^2w_x(\br_\alpha)  \,.
\label{rot inv eq vel gen}
 \end{eqnarray}
We see that, like the component $F_y$ of the force, the $y$-component of the dislocation velocity 
(which is, after all, proportional to $F_y$) vanishes for spatially uniform  $w_x$. And it does so for the same reason: a spatially uniform  $w_x$ corresponds to a uniform rotation, which cannot lead to dislocation motion in a rotation invariant system.

This result, based as it is purely on symmetry, proves to continue to apply even in active, rotation invariant smectics, as we will see in section [\ref{diseom}].

\subsection{Non-rotation invariant 2d equilibrium smectics: effects of a symmetry breaking field}\label{symbreakeq}
 
  \subsubsection{"Spin wave" (Phonon) theory with a symmetry breaking field: Quasi-long-ranged order}\label{sweqsymbreak}

As we have seen, rotation invariance plays a crucial role in the behavior of equilibrium two-dimensional smectics - indeed, in some sense, it makes the smectic phase impossible(at non-zero temperature)  in $d=2$. One can, however, make a 2d smectic non-rotation invariant. This can be done in a number of ways. Two of the simplest are:
 
  \noindent 1) applying a magnetic field (${\bf H}$) 
 
    \noindent 2) preparing the 2d surface on which the smectic lives in some non-rotation invariant way. For example, one could rub the surface in one direction with an abrasive cloth, or etch a set of parallel grooves along it.
    
Magnetic 
fields break rotation invariance by picking out a preferred direction for the layer normal 
\beq
\hn(\br)=-\sin[\phi(\br)]\,\hx+\cos[\phi(\br)]\,\hy \,,
\label{normaldef}
\eeq
where $\phi(\br)$ is the angle between the local layers at $\br$ and the $y$-axis.
They do this due to the fact that the magnetic susceptibility tensor $\chi^H_{ij}$ 
must, by symmetry, have the layer normal $\hn$ as one of their principal axes. This implies that the susceptibility tensor can be written in the form \cite{degennes, chaikin}
\beq
\chi^{H}_{ij}=\chi^{H}_0\delta_{ij}-\Delta\chi^{H}n_in_j \,,
\label{chiij}
\eeq
where the material parameters $\chi^{H}_0$ and $\Delta\chi^{H}$ are respectively the isotropic and anisotropic parts of the susceptibility.

The expression \ref{chiij}  in turn implies that the magnetic 
energy of the smectic are given by (note that the $H$ on the left-hand side of the following expressions stands for ``Hamiltonian", while the $H$ on the right-hand side of the first of them stands for magnetic field):
\beqn
H_{mag} &&={1\over2}\int d^2 r \chi^H_{ij}H_iH_j\nn\\&&=-{1\over2}\int d^2 r \Delta\chi^H ({\bf H}\cdot \hn)^2+{\rm constant} \,
\label{emag}
\eeqn
where the ``constant" 
includes those parts of the energy independent of the layer normal $\hn$.

 Inserting our expression \ref{normaldef}  for $\hn$, 
 and choosing our $y$-axis to be along the magnetic 
 field,   gives
 \beq
H_{mag} =-{1\over2}\int d^2 r \Delta\chi^HH^2\cos^2[\phi(\br)]+{\rm constant} \, .
\label{emagphi}
\eeq
Clearly, the magnetic 
energy favors (for positive $\Delta\chi^{H}$) alignment of $\hn$ along the 
magnetic fields (i.e., $\phi=0$). If, e.g., $\Delta\chi^{H}$) is negative, then the lowest energy configuration will have the layer normal $\hn$ {\it perpendicular} to ${\bf H}$. 
In these cases, we can still arrive at the expression \ref{emagphi} 
simply by choosing the $y$-axis to be perpendicular to the applied field. So the result \ref{emagphi} 
can easily be made to hold in general.
 
 Now assuming that fluctuations away from this minimum energy state are small, we can expand \ref{emagphi} 
 for small $\phi$, obtaining
 \beq
H_{mag} ={1\over2}\int d^2 r \Delta\chi^HH^2[\phi(\br)]^2+{\rm constant} \,
\label{ephismall}
\eeq
Finally, using the relation $\pp_xu=\phi$ (i.e.,   \eqref{rotation dif}) between $\phi$ and the layer displacement field $u$, we can rewrite this as
 \beq
H_{mag} ={1\over2}\int d^2 r \,g(\pp_xu)^2+{\rm constant} \,,
\label{esymbreak}
\eeq
where we've defined the ``symmetry breaking field strength" $g$ via
\beq
g\equiv \Delta\chi^HH^2
\label{gdefmag}
\eeq
for the case of an applied magnetic field.

Note that the symmetry breaking field strength $g$ is actually proportional to the {\it square} of the applied field. This is a consequence of the fact that the smectic is apolar; in a polar smectic, this symmetry breaking would be linear in the applied field, since the constituents of the smectic would then have spontaneous magnetic and electric
dipole moments.

The second approach mentioned above, of breaking symmetry by preparing the surface in a non-rotation invariant way, can be shown by similar arguments to lead to a symmetry breaking contribution to the Hamiltonian of the form \ref{esymbreak} as well. The dependence of $g$ on whatever quantity one uses to characterize the strength of the symmetry breaking of the surface preparation 
need not be quadratic, however. For example, if the preparation consisted of etching or rubbing a set of grooves onto the substrate, we would expect $g$ in  \ref{esymbreak} to be {\it linear}, not quadratic, in the density of such grooves, at least when that density is small.

Adding this additional symmetry breaking energy \ref{gdefmag} to the terms already present 
in our smectic energy gives the total smectic Hamiltonian for the non-rotation invariant case:
 \beq
H_{\rm sm}=\frac{1}{2} \int \dd^2r \left[B(\pp_y u)^2+g(\pp_x u)^2+K(\pp_x^2 u)^2\right]\, ,\label{Hsmsymbreak}
\eeq

The bend elasticity term $K(\pp_x^2 u)^2$ is clearly negligible, at long wavelengths (i.e., for small spatial gradients) relative to the symmetry breaking term $g(\pp_x u)^2$. We will therefore henceforth drop it, which leaves our Hamiltonian in the form:
    \beq
H_{\rm sm}=\frac{1}{2} \int \dd^2r \left[B(\pp_y u)^2+g(\pp_x u)^2\right]\, ,
\label{Hsmsymbreak2}
\eeq

This non-rotation invariant smectic problem can readily be seen to be equivalent to an XY model . To see this, make a simple linear change of variables from the layer spacing $u(\br,t)$ to an "angle field" $\theta(\br,t)$ defined via
\beq
\theta\equiv q_0 u\label{thetadef1}
\eeq
where $q_0\equiv{2\pi\over a_0}$ is the wavevector of the smectic layering. This has the effect of converting the invariance of the smectic system under the translation $u(\br, t)\to u(\br, t)+a_0$ (which is a symmetry since the smectic structure is periodic with period $a_0$ in the $y$-direction) to invariance under $\theta\to\theta+2\pi$. The latter symmetry implies that $\theta$ can be interpreted as the angle between a unit spin and some reference direction, or, equivalently, as the phase of a complex scalar. Both of these systems are XY models.

With the change of variables \eqref{thetadef1}, the Hamiltonian \eqref{Hsmsymbreak}
becomes
\beq
H_{XY}=\frac{1}{2} \int \dd^2r \left[K_y(\pp_y\theta)^2+K_x(\pp_x\theta)^2\right]\, ,\label{Hxy2}
\eeq
with 
\beq
K_y=Bq_0^{-2} \sep K_x=gq_0^{-2} \,.
\label{Kdef}
\eeq
 
We can convert this into the most familiar, isotropic form of the XY model by rescaling lengths anisotropically so as to make the coefficients of the two terms in 
\eqref{Hxy2} equal. It is easy to show that the change of variables 
\beq
x=x'\sqrt{K_x\over K_y}=x'\sqrt{g\over B}
\label{rescale}
\eeq
accomplishes this, leading to an isotropic model
\beq
H_{XYiso}=\frac{\bar{K}}{2} \int \dd x'\dd y |\nabla'\theta|^2\, ,\label{Hxy2}
\eeq
where $\nabla'\equiv\pp_{x'}\hx'+\pp_y\hy$, and the
spin wave stiffness $\bar{K}$ is just the geometric mean of $K_x$ and $K_y$:
\beq
\bar{K}=\sqrt{K_xK_y} \,.
\label{Kbar}
\eeq

 The model \eqref{Hxy2} with the ``compactness condition" that the operation $\theta\to\theta+2\pi$ takes  one to the same physical state is the extremely well-studied "XY model"\cite{KT}. It describes spin systems, with the local spin $\bS(\br')=(\cos(\theta(\br')), \sin(\theta(\br')))$, w here 
 \beq
 \br'\equiv (x', y)=\left(x\sqrt{B\over g} , y\right) \,.
 \label{r'def}
 \eeq
 It exhibits quasi-long-ranged order, that is, algebraically decaying spin correlations\cite{KT}:
\beq
\langle \bS(\br'_1)\cdot\bS(\br'_2)\rangle\propto |\br'_1-\br'_2|^{-\eta(T)} \,,
\label{qlroXY}
\eeq
with the non-universal, temperature-dependent exponent\cite{KT}
\beq
\eta(T)={\kbt\over2\pi \bar{K}}={\kbt\over2\pi \sqrt{K_xK_y}} ={\kbt q_0^2\over2\pi \sqrt{gB}} \,.
\label{etaxy}
\eeq

The correlation function in smectics most closely analogous to the spin-spin correlation function 
\eqref{qlroXY}  is the density-density correlation function. This follows from a standard result of the scattering theory of smectics \cite{degennes}, which states that the Fourier transformed equal-time correlations of the density near the $n$'th Bragg peak (i.e., near wavevector $\bq=nq_0\hy$ with integer $n$) are given by
\beq
\langle |\rho(\bq,t)|^2\rangle\propto\mathbf{FT}\bigg\{\langle \exp\{inq_0[u(\br'_1)-u(\br'_2)]\}\rangle\bigg\}\bigg |_{{\bf \delta q}} \,,
\label{rhocorrsmec}
\eeq
where $\mathbf{FT}$ denotes a Fourier transform and ${\bf \delta q}\equiv\bq-nq_0\hy$.
Mapping this to the XY problem by using the change of fields from $\theta$ to $u$ \eqref{thetadef1}, and the change of coordinates from $x$ to $x'$ \eqref{rescale}, we obtain power law correlations for the complex exponential in \eqref{rhocorrsmec}:
\beq
\langle \exp\{inq_0[u(\br'_1)-u(\br'_2)]\}\rangle\propto |\br'_1-\br'_2|^{-n^2\eta(T)} \,. 
\label{qlrosmectic}
\eeq
This is easily shown to imply that the Bragg peaks become power law singularities near the $n$'th Bragg peak:
\beq
\langle |\rho(\bq,t)|^2\rangle\propto |{\bf \delta q}|^{-2+n^2\eta(T)} \,.
\label{Braggeqsmec}
\eeq

This can be measured experimentally by various scattering techniques (either X-ray or light scattering, depending on the layer spacing $a$), or, in experiments in which the constituent particles can actually be imaged, by simply constructing the spatially Fourier transformed density correlations directly from particle positions.

\subsubsection{Equation of motion for the non-rotation invariant case}

Again taking the simplest, purely relaxational, equilibrium dynamics associated with this Hamiltonian, which is the \tdgl  EOM \eqref{TDGL}, we obtain

\begin{eqnarray}
\partial_t u = D_y \partial_y^2 u + D_x \pp_x^2 u  + f_u
\label{sw eom eq sym break}
\end{eqnarray}
where we've defined $D_y\equiv\Gamma B$ and $D_x\equiv \Gamma g$.

In (\ref{sw eom eq sym break}), we have, as in the rotation-invariant case, added a Gaussian white noise $f_u$ with correlations:
\begin{equation}
 \langle f_u ({\bf r},t)f_{u}({\bf r}',t')\rangle=2 \Gamma k_BT\delta(
 {\bf r}-{\bf r}') 
\delta(t-t')\,,
\label{sw noise sm}
\end{equation}
where the coefficient $\Gamma \kbt$ is required by the fluctuation-dissipation theorem\cite{chaikin}.

One somewhat surprising feature of the equation of motion \eqref{sw eom eq sym break}
is that, although it was derived from the {\it non}-rotation invariant free energy \eqref{Hsmsymbreak}, the equation of motion \eqref{sw eom eq sym break} itself {\it is} rotation invariant, as can be seen by noting that the equation remains unchanged under the substitution $u(\br, t)\to u(\br, t)+\phi x$, which corresponds to a uniform rotation of 
all 
 the smectic layers by an angle  $\phi\ll1$. We will use this observation later to argue that \eqref{sw eom eq sym break} is therefore the spin wave equation of motion we would expect even for  {\it rotation invariant active} smectics, for which there is no free energy.

  \subsubsection{Dislocation effects: Kosterlitz-Thouless transition}\label{disloeqsymbreak}
 
We can now treat dislocations in  smectics without rotation invariance exactly as we treated those with rotation invariance in the previous section. The definition of the field $\bw$ as the generalization of $\nabla u$ in the presence of dislocations is unchanged, as is the Burgers' condition \eqref{burgers dif w}. All that changes is the Hamiltonian, which is now \eqref{Hsmsymbreak} rather than \eqref{Hsmw rot inv w}. As a result, the Euler-Lagrange equation now becomes 
\beq
g\pp_xw_x(\br)+B\pp_yw_y(\br)=0 \,,
\label{ELGnon}
\eeq
which can be rewritten in Fourier space as 
\beq
gq_xw_x(\bq)+Bq_yw_y(\bq)=0 \,.
\label{ELGnonFT}
\eeq
Solving this simultaneously with the unchanged Burgers' condition \eqref{burgers dif w} gives
\beqn
&&w_x(\bq)={iBq_y\over gq_x^2+Bq_y^2}\sum_\alpha a_0 b_\alpha e^{i\bq\cdot\br_\alpha} \,,\\
&&w_y(\bq)={-igq_x\over gq_x^2+Bq_y^2}\sum_\alpha a_0 b_\alpha e^{i\bq\cdot\br_\alpha} \,.
\label{w solFT sym break}
\eeqn

Fourier transforming these solutions back to real space gives
\beqn
&&w_x(\br)=\sum_{\alpha} a_0 b_\alpha  G_x(\br-\br_\alpha) \,,\\
&&w_y(\br)=\sum_{\alpha} a_0 b_\alpha  G_y(\br-\br_\alpha) \,,
\label{w non rot inv}
\eeqn
where the Green's functions $G_{x,y}$ are now given by
\beqn
&&G_x(\br)=-{y\sqrt{gB}\over 2\pi(gy^2+Bx^2)}\,,\nonumber\\\\
&&G_y(\br)={x\sqrt{gB}\over  2\pi(gy^2+Bx^2)}\,.\nonumber\\
\label{G disloc sym break smec}
\eeqn

As we did for the rotation invariant smectic, we can calculate the energy of a dislocation configuration by inserting these results into the elastic Hamiltonian \eqref{Hsmsymbreak}, which can be rewritten in terms of the components of $\bw$ as
\beq
H_{\rm sm}=\frac{1}{2} \int \dd^2r \left[gw_x^2+Bw_y^2\right]\, .\label{Hsmw rsym break}
\eeq
Inserting our results \eqref{w rot inv}
and \eqref{G disloc sym break smec}
into this expression, and performing the integral over $\br$ gives
our dislocation Hamiltonian $H_{\rm disl}$ for non-rotation invariant smectics:

\beq
H_{\rm disl}=-{\sqrt{gB}\over 2\pi} \sum_{\langle\alpha\ne\beta\rangle} a_0^2 b_\alpha b_\beta\ln\left({|\br'_\alpha-\br'_\beta|\over a}\right) \,,
\label{Hdisloc aniso smec}
\eeq
where the sum is over pairs $\alpha$, $\beta$ of dislocations, with each pair counted once, and $\alpha\ne\beta$.
We remind the reader that $\br\equiv(x, y)$ and $\br'\equiv(x\sqrt{B\over g}, y)$.
Note that the sign of this expression implies that the potential between two oppositely charged 
dislocations is attractive.

From the form of this Hamiltonian, it is possible see why a Kosterlitz-Thouless defect unbinding transition must occur, and to even determine its temperature, by the following very simple argument, originally given (in a slightly different form) by Kosterlitz and Thouless\cite{KT}. 

Consider a minimal neutral pair of dislocations $\alpha=(1,2)$ of Burgers' charges:
$b_1=1$ and $b_2=-1$. From the dislocation Hamiltonian \eqref{Hdisloc aniso smec}, we see that  the energy of this pair will be
\beq
V(\bR)={\sqrt{gB}a_0^2\over2\pi}\ln\left({|\br'_1-\br'_2|\over a}\right)+2E_c \,,
\label{U disloc iso smec pair}
\eeq
where $\bR\equiv \br_1-\br_2$ is the  separation of the pair, and $E_c$ is a``core energy" that we have added to the energy of the pair to take into account the energy coming from distortions within a distance $a$ of the cores of the two dislocations, which are not accurately captured by our elastic theory, because that theory is only valid at long distances. This core energy $E_c$ also contains some constant ($\bR$-independent) contributions to the energy of the pair coming from the elastic energy outside this region. 
Hence, by simple Boltzmann statistics, the probability density $p(\bR)$ for this pair  is
\beqn
p(\bR)=\exp\left(-{V(\bR)\over\kbt}\right)=\kappa^2|R'|^{-\nu} \,,
\label{dislprob}
\eeqn
\vspace{.1in}

\noindent where we've defined the ``dislocation fugacity" $\kappa\equiv\exp(-E_c/\kbt)$, $\bR'\equiv(R_x\sqrt{B\over g}, R_y)$, and 
\beq
\nu\equiv{\sqrt{gB}a_0^2\over2\pi\kbt} \,.
\label{etaddef}
\eeq

The mean squared size of this dipole is
\bew
\beq
\langle|\bR|^2\rangle=\int d^2R \,p(\bR) |\bR|^2=\kappa^2\sqrt{g\over B}\int d^2R' \,|\bR'|^{2-\nu}\left({R\over R'}\right)^2 \,.
\label{msR}
\eeq
\ew
Since ${R\over R'}$ is bounded between $1$ and $\sqrt{B\over g}$, as the reader can easily convince herself, this mean squared value clearly diverges if $\nu\le4$. This signals dislocation unbinding; i.e., the Kosterlitz-Thouless transition, which corresponds to the loss of quasi-long-ranged translational order, or equivalently, the melting of the smectic into a nematic. This transition clearly occurs at the temperature $T_{KT}$ at which $\nu=4$; hence, using \eqref{etaddef}, we have
\beq
{\sqrt{gB}a_0^2\over2\pi k_BT_{KT}}=4 \,.
\label{TKT}
\eeq
Using this result, and  setting $T=T_{KT}$ in our expression \eqref{etaxy} for the exponent for algebraic decay of density correlations, we recover the famous result\cite{KT,NK}
\beq
\eta(T_{KT})=1/4 \,.
\label{eta14}
\eeq

The  crucial point the reader should take away from this discussion: first, is that, when the probability density for the separation of a dislocation pair falls off like a power law $R^{-\nu}$ with distance, dislocations will be bound if $\nu>4$, and unbound if $\nu<4$. We will later apply this criterion to active, non-rotation invariant  smectics to determine the critical noise strength at which dislocations unbind, melting the smectic phase.

Solving our expression \ref{TKT} for $T_{KT}$ gives
\beq
T_{KT}={\sqrt{gB}a_0^2\over8\pi k_B}\propto \sqrt{g} \,.
\label{TKTscale}
\eeq
Thus, we see that the smectic melting temperature grows quite rapidly with the symmetry breaking field strength $g$; specifically, like $\sqrt{g}$. We will see in section [\ref{dis sym break}] that, while {\it active} smectics can also be stabilized against dislocation unbinding by rotational symmetry breaking fields, they are much more delicate. Specifically, the critical noise strength (the analog in those non-equilibrium systems of the critical temperature) grows only linearly with the symmetry breaking field strength $g$.

Although we do not need to explicitly consider the dynamics of dislocations in order to understand their unbinding in this equilibrium problem, it is instructive to do so, in order to facilitate comparison with the active cases that we will consider next, for which the dynamics is all we have.

To study the dynamics, consider first
 a general pair of dislocations $\alpha=(1,2)$ of Burgers' charges:
$b_1$ and $b_2$, at positions $\br_1={\bf 0}$, $\br_2=\br=x\hx+y\hy$.
The dislocation Hamiltonian \eqref{Hdisloc aniso smec} then implies that the energy of this pair will be
\beq
V(\br)=-{\sqrt{gB} b_1b_2a_0^2 \over2\pi}\ln\left({|\br'|\over a}\right) \,.
\label{U disloc iso smec pair}
\eeq

The force experienced by dislocation  $\alpha=2$ will therefore be ${\bf F}=F_x\hx+F_y\hy$, with its Cartesian components $F_{x,y}$ given by
\bew
 \begin{eqnarray}
F_x&=&-\pp_xV(\br)={b_1b_2a_0^2 xB\sqrt{gB}\over  2\pi(gy^2+Bx^2)}=B a_0 b_2w_y^{(1)}(\br)
\nonumber\\
F_y&=&-\pp_yV(\br)
={b_1b_2a_0^2 yg\sqrt{gB}\over 2\pi(gy^2+Bx^2)}=- g a_0 b_2w_x^{(1)}(\br)
\label{sym break eq forces}
 \end{eqnarray}
\ew
where by $\bw^{1}(\br)$, we mean the contribution to the field $\bw$ at the position $\br$ of the $\alpha=2$ dislocation coming from the $\alpha=1$ dislocation (i.e., neglecting the field created by dislocation $\alpha=2$ itself). It is straightforward to show that the generalization of these forces to configurations with more than two dislocations is:
 \begin{eqnarray}
F_x^\alpha&=&- B a_0 b_\alpha w_y(\br_\alpha)
\nonumber\\
F_y^\alpha&=&g a_0 b_\alpha w_x(\br_\alpha)
\label{sym break eq forces gen}
 \end{eqnarray}
where by $\bw(\br)$, we mean the contribution to the field $\bw$ at the position $\br_\alpha$ of the $\alpha$'th dislocation coming from all of the other dislocations (i.e., neglecting the field created by dislocation $\alpha$ itself).

Since the dislocation cannot "tell" whether the local field $\bw$ is created by other dislocations, by spin waves, or by externally applied stresses, we expect \eqref{rot inv eq forces gen} to hold more generally if we take $\bw_\alpha$ on the right-hand side of those equations to be the entire $\bw$ field, {\it excluding} the part due to dislocation $\alpha$ itself.
 This proves to be important when we consider the effect of stresses at the boundary on dislocation   motion.

There are two important features of the result \eqref{rot inv eq forces gen} that should be noted:

\noindent 1) the force on a given dislocation $\alpha$ is determined entirely by the {\it local} value of the field $\bw(\br_\alpha)$ (and its derivatives) at the location $\br_\alpha$ of that dislocation  (excluding the part of that field due to the given dislocation itself).

\noindent 2) The force in this non-rotation invariant case now depends directly on the $x$ component $w_x$; no spatial derivatives are required. This means in particular that a uniform $w_x$ {\it does} generate a force on the dislocation. This is a consequence of the {\it lack} of rotation invariance: as shown by equation \eqref{rotation dif}, a spatially uniform $w_x$ corresponds to a uniform rotation of the layers, which now {\it can} lead to a force on the dislocation, since the system is {\it not}   rotation-invariant. This consideration will continue to apply in active smectics, and will allow certain terms in the force in those systems in the rotation {\it non-invariant} case which are absent in the rotation invariant case.

As in the rotation-invariant case, we expect the velocity $\bv$ of the dislocations to be linear in the
force $\bF$. That is,
\beq
\bv_\alpha=\bmu\bF_\alpha \,,
\label{mobtens}
\eeq
where $\bmu$ is a constant "mobility tensor". On the same symmetry grounds as before, we expect this tensor to be diagonal in our $(x,y)$ coordinate system (i.e., with the $x$ and $y$ axes respectively parallel
and perpendicular to the mean smectic layers); hence
\beq
v_x^\alpha=\mu_x F_x^\alpha \,, \, v_y^\alpha=\mu_y F_y^\alpha \,.
\label{vdisiso}
\eeq
Using our earlier results \eqref{sym break eq forces gen} for the forces on the dislocations, we can rewrite these as
 \begin{eqnarray}
v_x^\alpha&=&-  \mu_x B a_0 b_\alpha w_y(\br)
\nonumber\\
v_y^\alpha&=& \mu_y g a_0 b_\alpha w_x(\br)  \,.
\label{non rot inv eq forces gen}
 \end{eqnarray}
We see that, like the component $F_x$ of the force, the $x$-component of the dislocation velocity
(which is, after all, proportional to $F_x$) no longer vanishes for spatially uniform  $w_x$. And it need not, since, although  a spatially uniform  $w_x$ still corresponds to a uniform rotation,  in a {\it non}-rotation invariant system, no symmetry forbids a uniform rotation from causing  dislocation motion.

\subsection{Summary of the equilibrium cases}

We have seen that, in equilibrium, rotation invariant  2d smectics, translational order at non-zero temperature is always short ranged, even in spin wave theory (i.,e., when dislocations are ignored). Furthermore, dislocations are always unbound at any non-zero temperature, if the smectic is rotation invariant. Effectively, this means that 2d smectics melt as soon as the temperature becomes non-zero. Another way to say this is that 2d smectics do not exist at temperatures $T>0$.

Breaking rotation invariance by, e.g., applying a rotational symmetry breaking magnetic 
field,  or breaking the underlying rotation invariance in other ways, can stabilize {\it quasi}-long-ranged translational order (i.e., power law decay of translational correlations) in spin wave theory. Rotational symmetry breaking also stabilizes two-dimensional equilibrium smectics against dislocation
unbinding. The temperature $T_m$ at which these systems melt vanishes as the strength $g$ of the applied symmetry breaking field vanishes, according to the law \ref{TKTscale}.

We will see that, while the presence or absence of rotation invariance has no important effect on the spin wave dynamics of active smectics, nor on the fields created by dislocations, it has a profound effect on the motion of dislocations. In fact, like equilibrium 2d smectics,  active 2d smectics are only stable against dislocations if rotation invariance has been explicitly broken by an externally applied symmetry breaking field.  Furthermore, the field required to stabilize active 2d smectics against dislocations is much higher than the field needed in equilibrium 2d smectics. 

 \section{"Spin wave" (Phonon) theory of active smectics}\label{swact}
 
 \subsection{Rotation invariant 2d active smectics: spin wave theory}\label{rotinvact}
 
In this section, we will review the hydrodynamic theory of active, rotation-invariant, apolar smectics. For more details, the interested reader is referred to \cite{apolarsm}. We will first limit our discussion to ``spin-wave theory"; that is, dislocation-free smectics.

Because there are no conserved quantities, the only hydrodynamic field in our problem is the layer displacement $u$, which is the "Goldstone mode" associated with the breaking of translational symmetry by the layering.  The  
long-wavelength hydrodynamics of this field
is therefore simply the most general equation of motion, to leading order in space and time derivatives, that respects the symmetries of this system. These symmetries are rotation and translation invariance. As we noted earlier in our discussion of the {\it non-rotation invariant equilibrium} smectic,
the equation of motion \eqref{sw eom eq sym break} for that system, oddly, {\it is} rotation invariant. Therefore, that equation, which we repeat here for the readers' convenience, also describes {\it active, rotation invariant} smectics:
\begin{eqnarray}
\partial_t u = D_y \partial_y^2 u + D_x \pp_x^2 u  + f, \label{layerdyneq}
\label{sweom}
\end{eqnarray}
where $f^u$ is a Gaussian, zero-mean spatiotemporally white noise with 
\begin{equation}
 \langle f_u ({\bf r},t)f_{u}({\bf r}',t')\rangle=2 \Delta\delta(
 {\bf r}-{\bf r}') 
\delta(t-t')\,.
\label{sw noise sm}
\end{equation}
Because this is a non-equilibrium system, there is no longer a fluctuation-dissipation theorem\cite{chaikin} relating the noise variance $\Delta$ to the dissipative terms in \eqref{sweom}. However, the equation of motion \eqref{sweom}, with the noise correlations \eqref{sw noise sm} is, as noted, identical to that of an equilibrium, non-rotation invariant smectic with 
\beq
\Gamma\kbt=\Delta \, \sep \Gamma B=D_y \,\sep \Gamma g=D_x \,.
\label{equivs}
\eeq
We can therefore use these relations to obtain any spin-wave correlation function in the active, rotation-invariant smectic from the corresponding correlation function in an equilibrium, non-rotation invariant smectic.
In particular, this reasoning predicts , {\it in the absence of dislocations}, that active, rotation invariant smectics will exhibit 
power law singularities near the $n$'th Bragg peak (i.e., for wavevector $\bq=nq_0\hy+\delta\bq$ with integer $n$ and $|\delta\bq|\ll q_0)$:
\beq
\langle |\rho(\bq,t)|^2\rangle\propto |{\bf \delta q}|^{-2+n^2\eta(\Delta)} \,,
\label{Braggeqsmecactive}
\eeq
with the non-universal exponent\cite{KT}
\beq
\eta(D_x, D_y, \Delta)={\Delta q_0^2\over2\pi \sqrt{D_xD_y}} \,.
\label{etaxyactive}
\eeq

As in equilibrium systems, this can be measured experimentally by various scattering techniques (either X-ray or light scattering, depending on the layer spacing $a$), or, in experiments in which the constituent particles can actually be imaged, by simply constructing the spatially Fourier transformed density correlations directly from particle positions. It can be determined from simulations by the latter approach as well.

Interestingly, as we shall see, in rotation invariant active smectics, this quasi-long-ranged order is destroyed by unbound dislocations, 
and they therefore do not exhibit any singular behavior at the Bragg spots at all. In rotation {\it non-invariant} active smectics, however, the results \ref{Braggeqsmecactive}
and \ref{etaxyactive}
will hold for sufficiently small noise.

Clearly,  the non-universal exponent $\eta(D_x,D_y, \Delta)$ for algebraic decay of translational correlations, when expressed in terms of $D_x$ and $D_y$ will continue to be given in terms of $D_x$, $D_y$, and the noise strength $\Delta$ (which we also expect to be independent of the symmetry breaking field $g$ for small $g$) by the result found earlier for rotation invariant active smectics; i.e.,

\beq
\eta(T)={\kbt q_0^2\over2\pi \sqrt{gB}}={\Delta q_0^2\over2\pi \sqrt{D_xD_y}} \,.
\label{etaxyactive}
\eeq

 Thus, it would {\it appear}, from this spin wave theory, that active smectics are more robust against fluctuations that equilibrium ones: it looks like we can get quasi-long-ranged translational order in these systems even {\it without} breaking 
rotation invariance. As we'll see in section \eqref{dis unbind}, this conclusion is actually wrong, due to the unbinding of dislocations.

One last comment about our equation of motion \eqref{sweom} for active, rotation-invariant smectics is in order. The term  $D_x\pp_x^2u$  \cite{actten,Ramaswamy2000}
is, as we discussed in section [\ref{rotinveq}], forbidden
in \textit{equilibrium} rotation-invariant smectic, where it would correspond to a term $\propto(\pp_xu)^2$ in the Hamiltonian \eqref{Hsm rot inv} 
, which is forbidden by rotation invariance.  In our out-of-equilibrium system, however, only the
equation of motion itself must be rotation-invariant, which we have already shown equation \ref{sweom} is.  Physically, the origin of this term is that local vectorial asymmetry of a curved layer must inevitably
lead to directed motion in a self-driven system.

 \subsection{Non-rotation invariant 2d  active smectics: effects of a symmetry breaking field (spin wave theory)}\label{symbreakact}

We now turn to the spin-wave theory for active smectics in the presence of a symmetry breaking field. Even when rotational symmetry is broken, the lowest order in derivative terms allowed by $x\to-x$, $y\to-y$ symmetry are still second order derivatives. Therefore, the equation of motion is still:
\begin{eqnarray}
\partial_t u = D_y \partial_y^2 u + D_x \pp_x^2 u  + f,
\label{sweomsymbreak}
\end{eqnarray}
where $f^u$ is a Gaussian, zero-mean spatiotemporally white noise with
\begin{equation}
 \langle f_u ({\bf r},t)f_{u}({\bf r}',t')\rangle=2 \Delta\delta(
 {\bf r}-{\bf r}')
\delta(t-t')\,,
\label{sw noise sm sym break}
\end{equation}
as in the rotation-invariant case.
The only difference between this and the equilibrium case is that, in equilibrium, $D_x$ can only be non-zero due to the presence of the symmetry breaking field. Indeed, we expect $D_x\propto g$. In active smectics, on the other hand, as we've just seen, $D_x$ can be non-zero simply due to the activity.
Thus, for small symmetry breaking field $g$ (the limit we will consider later), we expect $D_x$ to be essentially independent of the symmetry breaking field $g$.

\section{Dislocations in active smectics: configurations}\label{disconf}

In our non-equilibrium system, we can no longer determine the fields of the dislocations by minimizing the free energy, since there is no  free energy for a non-equilibrium system. However, we can readily obtain the fields of static dislocations simply by looking for steady state solutions of the equations of motion \eqref{sweom}, once those equations are suitably rewritten to take into account the presence of dislocations. As has already been discussed, this amounts to making the replacements $\pp_xu\to w_x$ and $\pp_yu\to w_y$. Doing so in the equation of motion \eqref{sweom}, and setting all time derivatives to zero, gives
\beq
D_x\partial_xw_x(\br)+D_y\pp_yw_y(\br)=0 \,.
\label{steady state}
\eeq

The other condition on dislocations is the Burger's condition, which can be written  
\beq
\pp_xw_y(\br)-\pp_yw_x(\br)=\sum_\alpha a_0 b_\alpha\delta(\br-\br_\alpha) \,.
\label{Burger}
\eeq

These two simultaneous linear equations \eqref{steady state FT} and \eqref{BurgerFT} are exactly the same as those we obtained for {\it non-rotation-invariant, equilibrium} smectics if we make the identifications \eqref{equivs}. Therefore, we can simply transcribe the solutions we obtained for that problem here. This gives
\beqn
&&w_x(\br)=\sum_{\alpha} a_0 b_\alpha  G_x(\br-\br_\alpha) \,,\\
&&w_y(\br)=\sum_{\alpha} a_0 b_\alpha  G_y(\br-\br_\alpha) \,,
\label{w rot inv act}
\eeqn
where the Green's functions $G_{x,y}$ are now given by
\beqn
&&G_x(\br)=-{y\sqrt{D_xD_y}\over 2\pi(D_xy^2+D_yx^2)}\,,\nonumber\\\\
&&G_y(\br)={x\sqrt{D_xD_y}\over  2\pi(D_xy^2+D_yx^2)}\,.\nonumber\\
\label{G disloc rot inv act}
\eeqn

Note that we can write these Greens functions in terms of the gradient of a potential:
 \beqn
 G_x(\br)=-\varpi\pp_yV(\br) \,,\nonumber\\
G_y(\br)={1\over\varpi}\pp_xV(\br) \,,
\label{rot inv fields}
 \eeqn
where the potential
\beq
V(\br)={1\over2\pi}\ln\left({|\br'|\over a_0}\right)={1\over4\pi}\ln\left({D_yx^2+D_xy^2\over D_xa_0^2}\right) \,,
\label{U active}
\eeq
and we've defined $\varpi\equiv\sqrt{D_y\over D_x}$ and $\br'\equiv(x\sqrt{D_y\over D_x}, y)$.

These dislocation fields are essentially identical in form to those we found
for equilibrium, {\it non-rotation invariant} smectics in section [\ref{symbreakeq}]. However, we will see in the next section that the {\it motion} of dislocations in response to these fields is very different from that case.

\section{Dislocation equation of motion for rotation invariant active smectics}\label{diseom} 


We have established that both the spin wave theory of active, rotation-invariant smectics, and the
field $\bw(\br)$ generated by dislocations, are the same as those found for a {\it non-rotation invariant} equilibrium smectic. It might therefore seem reasonable to assume that the motion of dislocations, and, as a result, the dislocation unbinding transition in these active, rotation invariant systems would be the same as those of the equilibrium, non-rotation invariant system. Indeed, precisely this argument was made in earlier publications by one of us\cite{apolarsm, polarsm}.

However, this conclusion proves to be wrong. The reason is that, despite the just noted similarities  between active rotation invariant systems and equilibrium, non-rotation invariant systems, active, rotation invariant smectics are still rotation invariant. This fairly obvious (indeed, tautological!) statement makes the motion of dislocations in an active, rotation-invariant smectic very different from that in an equilibrium, non-rotation invariant one. The motion is so different, in fact, that although dislocations are always bound at sufficiently low temperatures in equilibrium, non-rotation invariant  smectics at sufficiently low temperature, they are {\it never} bound  in an active, rotation-invariant smectic with {\it any} non-zero noise, no matter how small.

We will now demonstrate this. We begin by deriving the equation of motion for dislocations in an active, rotation-invariant smectic.

We restrict ourselves to
``unit" dislocations,  by which we mean a dislocation whose Burgers' number $b$ has the smallest possible magnitude; i.e., $b=\pm 1$.

As can be seen from figure [\ref{Burgers'}], and from our analytic solutions \eqref{w rot inv act} for the dislocation fields, 
a dislocation in a smectic is an inherently {\it polar} object; it breaks the left-right symmetry along the layers. This means that, in an active system, there is no symmetry forbidding spontaneous motion of dislocations either left or right along the layers. Therefore, by the Curie principle \cite{curie},  that ``anything that's not forbidden is compulsory", we expect such motion to occur.

Since a unit dislocation with $b=-1$ is just the mirror image of one with $b=1$, if $b=+1$ dislocations spontaneously move to the left, $b=-1$ must spontaneously move to the right, and visa-versa. The motion should be along the local layer direction, since spontaneous motion {\it perpendicular} to the layers is forbidden by the fact that dislocations do {\it not} break {\it up-down} symmetry.

Of course, a local {\it curvature} of the layers will break up-down symmetry. Indeed, this is the origin of the force in equilibrium smectics in the $y$-direction given in equation \eqref{rot inv eq forces gen}. Note, however, that because this effect involves {\it curvature}, it involves at least two derivatives of the displacement field, or, equivalently, one derivative of $\bw$. Hence, to leading (i.e., zeroth) order in derivatives of $\bw$, the motion of a dislocation must be {\it along} the layers. This has profound implications for the stability of the active smectic state in rotation invariant systems, as we will see.

These considerations imply that, to zeroth order in gradients of $\bw$, the velocity $\bv_\alpha$ of the $\alpha$'th dislocation  must take the form:
\beq
\bv_\alpha=v_s{\rm sgn}(b_\alpha)\hz\times\hn(\br_\alpha)
\label{selfp1}
\eeq
where 
\beq
\hn\approx\hy-\phi(\br)\hx
\label{normal}
\eeq
is the local normal to the smectic layers. The characteristic speed $v_s$ appearing in \eqref{selfp1} is a system-dependent parameter. It will depend, importantly, on local properties like the mean layer spacing $a$ of the active smectic. It is this dependence that makes it possible for the active smectic to reach homeostasis, as we'll argue below.

The {\it direction} $\hz\times\hn$ of $\bv$ is dictated by the requirement that this 
spontaneous motion be {\it along} the layers, which, as we just discussed, is required by up-down symmetry. The factor of ${\rm sgn}(b_\alpha)$ simply reflects the fact noted above that oppositely charged dislocations must move in opposite directions.

Since our definition \eqref{wdef} of $\bw$ implies that the local layer spacing is simply 
\beq
a(\br)=a_0(1+w_y(\br)) \,,
\label{a(r)}
\eeq
where $a_0$ is the ``reference" layer spacing , relative to which we measure $\bw$, the dependence of the spontaneous speed $v_s$ on the layer spacing $a$ is equivalent to dependence on $w_y$. That is, we can (and will!) take $v_s$ to be a local function of $w_y$.

Note that rotation invariance forbids any dependence of $v_s$ on $w_x$, since a uniform $w_x$ corresponds to a pure rotation, which cannot change the spontaneous speed (or, indeed, any local scalar) in a rotation invariant system.

These arguments imply that the spontaneous speed $v_s(\br)$ at a point $\br$  can be written as
\beq
v_s(\br)=v_s(a(\br))=v_s(a_0(1+w_y(\br))\equiv v_s(w_y(\br)) \,.
\label{v0wy}
\eeq

Using \eqref{v0wy} in our expression \eqref{selfp1} for the dislocation velocity gives
 \begin{eqnarray}
v_x^\alpha&=&v_s{\rm sgn}(b_\alpha)+\mu_x b_\alpha w_y(\br_\alpha)
\nonumber\\
v_y^\alpha&=&v_s  {\rm sgn}(b_\alpha)w_x(\br_\alpha)  \,.
\label{rot inv active forces gen}
 \end{eqnarray}
 
 We will argue in the next section that, for an active smectic confined between fixed boundaries,
 $v_s$
vanishes in the steady state. We will refer to this state as the state of ``homeostasis". Note that this implies that there is no motion in the $y$-direction (i.e., normal to the smectic layers) in the homeostatic state. This in turn implies that, in the presence of noise, dislocations in an active smectic are always unbound. As a result,  the active smectic state is, at {\it any} non-zero noise, always destroyed by dislocation unbinding in rotation invariant active smectics.

\section{Dislocations: Self-propulsion and the approach to homeostasis}\label{homeo}

The dominant term in the equations of motion \eqref{rot inv active forces gen} is the ``self-propulsion" term $v_s{\rm sgn}(b_\alpha)$. For $v_s>0$, this term will make positive dislocations move to the right, and negative dislocations move to the left. Obviously, this switches for $v_s<0$.

Because $w_{x,y}(\br_\alpha) \to0$ as $|\br|\to\infty$ (as can be seen from equation \eqref{w rot inv act}), the interactions between dislocations cannot compete with this constant ``external force" -like motion. Hence, even tightly bound pairs of dislocations will eventually be rent asunder by the spontaneous motion , with all of the positive dislocations moving to one side, and all of the negative dislocations moving to the other.

We therefore expect dislocation pairs to constantly nucleate, be ripped apart by this spontaneous velocity, and traverse the system.

Consider first the case $v_s(w_y=0)>0$. In this case, all positive dislocations will eventually traverse the system from left to right, while all negative dislocations will eventually traverse the system from right to left. It is easy to see, both by inspection of figure \ref{Burgers'}, and from our expressions  for the dislocation fields, that each time one of these happens (i.e., each time either a positive dislocation traverses the system from left to right, or a negative dislocation traverses the system from right to left), the number of layers in the system is increased by one. Therefore, if the boundaries of the system at the top and bottom remain fixed, the strain $w_y$ will decrease (i.e., become more negative), since the mean layer spacing is reduced.

The crucial point here is that this self-propelled dislocation motion changes the mean strain $w_y$ in the system. But since the dislocation self-propulsion speed $v_s(w_y)$ is itself a function of $w_y$, this implies that, as this process of pair nucleation, separation, and motion across the system continues, the strain $w_y$ will continue to evolve.

Will it ever stop? Yes, it will, provided that the speed $v_s(w_y)$ vanishes at some negative $w_y$. We would expect it to do so: as shown by \eqref{non rot inv eq forces gen}, in an equilibrium smectic, a more negative $w_y$ causes positive dislocations to move to the left dislocation with a speed proportional to $w_y$. 
Stability requires
that the negative strain induced by the process just described will oppose the motion of positive dislocations to the right. This opposition should get stronger as $w_y$ becomes more negative, so it is very plausible that a type of ``homeostasis"  will eventually be reached, at which the strain $w_y$ takes on a value  $w_{y, {\rm h}}$ such that
\beq
v_s(w_{y, {\rm h}})=0 \,.
\label{homeo}
\eeq

This will almost always happen whenever the direction of spontaneous motion of the dislocations causes them to move in a direction that makes the strain evolve in such a way as to oppose the spontaneous motion of the dislocations. As just discussed, this is most likely to occur if $v_s(w_y=0)>0$, and 
${dv_s\over dw_y}>0$ giving rise to a steady state extensile stress normal to the layers. The opposite case of $v_s(w_y=0)<0$, and ${dv_s\over dw_y}<0$ will also reach homeostasis, and give rise to a contractile steady state stress normal to the layers.

So we expect stable
active smectic systems to reach homeostasis, as defined by \ref{homeo}. In such systems, the homeostatic layer spacing $a_{h}$ will be
\beq
a_{h}=a_0(1+w_{y, {\rm h}}) \,.
\label{ahom}
\eeq
It clearly makes sense to define this homeostatic layer spacing as our reference layer spacing, and measure our displacement field $u$ relative to that. It is easy to relate the $u$ field $u_{h}(\br)$ defined relative to this homeostatic state to that $u_0(\br)$ defined relative to the initial state with layer spacing $a_0$; the required transformation is just a linear function of the Cartesian coordinate $y$:
\beq
u_{h}(\br)=u_0(\br)-w_{y, {\rm h}}y \,.
\label{uhome}
\eeq
Henceforth, we will drop the subscript ${\rm h}$, and implicitly assume that our $u$ field is measured relative to the homeostatic state. With this choice of variables, the homeostatic condition \ref{homeo} becomes simply
\beq
v_s(w_y=0)=0 \,.
\label{homeoref}
\eeq

Note that the argument just presented for the approach of active smectics to the
homeostatic state depended on the presence of {\it fixed} boundaries without layer flux
at the top and bottom of the system.
For active smectics confined under constant {\it stress} between movable boundaries, however, a homeostatic state is never reached. Instead, the active smectic either grows arbitrarily large, pushing the boundaries ever further out, or shrinks and disappears. The unbounded growth scenario occurs if the applied normal stress $\sigma_n$ at the boundaries is less than some homeostatic value $\sigma_c$, while shrinkage and disappearance occurs if $\sigma_n>\sigma_c$.

The reason for this behavior is clear: like all elastic systems, fixing the stress is equivalent to fixing the strain. In smectics, the ``strain" is just $w_y$. Hence, by fixing the stress at the boundary, we fix $w_y$. If $w_y$ is fixed at a value such that $v_s(w_y)>0$ (this corresponds to $\sigma_n<\sigma_c$), then positive dislocations will move to the right, and negative ones to the left, thereby adding layers to the system. The only way to keep $w_y$ fixed, therefore, is for the top and bottom surfaces to move out, to accommodate the extra layers. This process will continue indefinitely.

On the other hand, if $w_y$ is fixed at a value such that $v_s(w_y)<0$ (this corresponds to $\sigma_n>\sigma_c$), then positive dislocations will move to the right, and negative ones to the left, thereby adding layers to the system. The only way to keep $w_y$ fixed, therefore, is for the top and bottom surfaces to move in, to accommodate the loss of layers. This process will continue until the active smectic disappears completely.
This behavior is reminiscent of that predicted for tissues  and observed in epithelia \cite{homeoTissue},\cite{homeoTissuetwo}, \cite{homeoTissueexp}. The addition and removal of layers corresponds to
the addition of cells by cell division and the removal of cells by cell death. In a tissue under homeostatic conditions, the cell division rate is exactly balanced by cell death rate. Since both cell division and cell death depend on tissue pressure, in general there will be only one tissue pressure value for which these rates balance exactly. This defines the homeostatic pressure, corresponding to the stress $\sigma_c$ in the homeostatic state of the smectic. If the tissue is given a prescribed volume, keeping constant biochemical conditions, cells will divide and the tissue will grow to occupy all space and settle to steady state, ie homeostasis, when the tissue pressure reaches the homeostatic pressure. Alternatively if the tissue is kept at a constant pressure larger than the homeostatic one, cell death will win over cell division and the tissue will disappear. If the pressure is kept at a lower value the tissue will invade all space. Tissue invasion of one type of tissue by another one, will occur if the homeostatic pressure of the invad{\it ing}
 tissue  is larger than the homeostatic pressure of the invad{\it ed} tissue.

  While the above discussion has been very physical, biochemistry nonetheless plays an important role, since the homeostatic pressure values depend on local biochemical conditions.

Returning now to the case of fixed boundaries, and the resultant homeostatic state, we can now ask 
 , since the spontaneous velocity tearing dislocation pairs apart vanishes in the homeostatic state, whether or not it is possible to achieve a state free of unbound dislocations. Only if such a state is possible can we have a true smectic.

In the presence of noise, the answer to this question proves to be no: dislocations in a rotation invariant active smectic will always be unbound. We will demonstrate this in the next section.

\section{Motion at homeostasis and the destruction of the active smectic phase}\label{dis unbind} 

As shown in the last section, at homeostasis, 
\beq
v_s(\br)=v_s(a(\br))=v_s(a_{\rm h}(1+w_y(\br))\approx \mu_x w_y(\br) \,,
\label{v0expand}
\eeq
where in the last, approximate, equality, we have expanded for small $w_y(\br)$, and defined the ``mobility" $\mu_x\equiv {1\over a}
\left({dv_s(a)\over da}\right)_{a=a_{\rm h}}$. We have also used the fact that 
\beq
v_s(a_{\rm h})=0 \,,
\label{v0hom}
\eeq
since, as discussed above, isolated dislocations do not spontaneously move in the homeostatic state.

Inserting \ref{v0expand} into our general equation of motion \eqref{rot inv active forces gen} for the dislocations, and linearizing those equations in the strain field $\bw$ gives
 \begin{eqnarray}
v_x^\alpha(t)&=&\mu_x b_\alpha w_y(\br_\alpha,t)+f^\alpha_x(t) \,,
\nonumber\\
v_y^\alpha&=&f^\alpha_y(t) \,,
\label{rot inv active forces hom}
 \end{eqnarray}
 where we have added a ``Langevin force" - i.e., a random white noise $\bff^\alpha$ - to the equation of motion.  In equilibrium, these would simply be thermal noises, with variances proportional to temperature. In our non-equilibrium system, they can have non-thermal, active contributions as well. We will assume, as seems reasonable, that these forces are white, Gaussian, zero-mean, and decorrelated between different dislocations. Taking these conditions together with the $x\to-x$, $y\to-y$ symmetries of the apolar smectic state we're considering in this paper,  implies that these forces have correlations
 \beqn
 \la\bff^\alpha(t)\ra&&={\bf 0} \,\nn,\\
 \la f^\alpha_i(t) f^\beta_j(t') \ra&&=\Bigg(\Delta_x \delta_{ij}^x+\Delta_y \delta_{ij}^y\Bigg)\delta(t-t')\delta_{\alpha\beta}\,.\nn\\
\label{disforcecorr}
 \eeqn
Here $\delta_{ij}^k=1$, when $i=j=k$ and $0$ otherwise.
Since the forces are assumed to be Gaussian random variables (as suggested by the central limit theorem), these correlations completely specify the distribution of the forces $\bff^\alpha(t)$.

This implies that, to linear order in the field $\bw$, dislocation motion in the $y$-direction is a perfectly random walk. Therefore, any pair of dislocations will eventually wander arbitrarily far apart in the $y$-direction. Hence, dislocations in active, rotation invariant smectics are always unbound. Another way to say this is that a true active smectic phase cannot exist in a noisy, rotation invariant system. 

\vspace{.2in}

\section{Binding dislocations with a symmetry breaking field}\label{dis sym break} 

As in equilibrium smectics, it is possible to make dislocations bind in 2d active smectics by applying symmetry breaking fields. Once we do so, there is no longer any symmetry argument forcing the dislocation velocity $\bv_s$ to be independent of the ``rotational" component of the strain $w_x$. Nor must this velocity be directed along the layers. Therefore, we can write in general:
\beq
\bv_s^\alpha=\bigg(v_{s\parallel}\hz\times\hn(\br_\alpha)+v_{s\perp}\hn(\br_\alpha)\bigg){\rm sgn}(b_\alpha)
\label{selfp2}
\eeq
 
By up-down symmetry, $v_{s\perp}(w_x=0)=0$. Therefore, the leading order term in the expansion of $v_{s\perp}(w_x)$ in powers of $w_x$ is the linear term:
 \beq
 v_{s\perp}(w_x)=-\mu_yw_x \,,
 \label{vdperpexp}
 \eeq
where we've defined
\beq
\mu_y\equiv-\left({dv_{d\perp}(w_x)\over dw_x}\right)_{w_x=0} \,.
\label{muydef}
\eeq

By the same reasoning about homeostasis that we applied to the rotation invariant case, we expect that, at homeostasis,
\beq
v_{s\parallel}(a_{\rm h})=0 \,.
\label{vparhom}
\eeq
We can therefore again choose the state with $a=a_{\rm h}$ to be our reference state, and expand for small strain $w_y$, obtaining, just as we did for the rotation invariant case,
\beq
v_{s\parallel}(\br)=v_s(a(\br))=v_s(a_{\rm h}(1+w_y(\br))\approx \mu_x w_y(\br) \,,
\label{vdparexp}
\eeq
where again, we have excluded a term proportional to $w_x$ by up-down symmetry.

Inserting the results   (\ref{vdperpexp}) and  (\ref{vdparexp})  into  (\ref{selfp2}), using again the relation \ref{normal} $\hn\approx\hy-\phi(\br)\hx$ between the layer normal $\hn$ and the strain $w_x$,
and expanding to linear order in the strain $\bw$, we obtain
\beq
v_x=\mu_x w_y \,, \, v_y=-\mu_y w_x \,.
\label{vdis}
\eeq

Since the $\mu_y$ term in \eqref{vdis} only appears due to the rotational symmetry breaking field, $\mu_y$ must vanish as the strength $g$ of that symmetry breaking field goes to zero. We therefore expect
\beq
\mu_y\propto g \,
\label{muyscale}
\eeq
for small symmetry breaking field strength $g$.
This is the {\it only} parameter in the dislocation dynamics that
we expect to exhibit any strong dependence on $g$ at small $g$. We will use this fact later to determine the behavior of the critical noise strength at which dislocation unbinding occurs with $g$.

Note also that, as we saw in our discussion in section [\ref{sweqsymbreak}] of symmetry breaking fields in {\it equilibrium} smectics, $g$ could scale non-linearly with experimentally tunable parameters like magnetic field $H$. 
 Indeed, we expect that, as in equilibrium, $g\propto H^2$ 
 for magnetic 
 symmetry breaking fields. On the other hand, for symmetry breaking induced by etching grooves into the 2d surface, we expect $g$ to scale linearly with the density of those grooves.

Consider now an isolated  neutral pair of fundamental dislocations, one with Burgers number $b_+=+a$ located at $\br_+\equiv(x_+, y_+)$, the other with Burgers number $b_-=-a$ located at $\br_-\equiv(x_-, y_-)$.

Since, as discussed earlier, the spin-wave equation of motion \eqref{sweomsymbreak} is unchanged in form by the presence of the symmetry breaking fields, and since, furthermore, the Burgers' condition is also unchanged, we can use the expressions \eqref{w rot inv act} for the dislocation fields for active, {\it rotation invariant} active smectics for this {\it non-rotation invariant} case as well. Therefore, 
the strain field $\bw(\br_+)$ due to the $-$ dislocation at the position $\br_+$ of the $+$ dislocation is, from equation \eqref{w rot inv act},
\beq
w_x(\br_+)=-a G_x(\br) \, \sep w_y(\br_+)=-aG_y(\br)  \,,
\label{+field}
\eeq
where we've defined the relative displacement $\br\equiv(x_+-x_-, y_+-y_-)\equiv(x,y)$. As always, we exclude the field of the $+$ dislocation at $\br_+$ from the field $\bw$ above, since the $+$ dislocation only responds  to the field of the {\it other} dislocation.
Likewise, the strain field $\bw(\br_-)$ due to the $+$ dislocation at the position $\br_-$ of the $-$ dislocation is, from equation \eqref{w rot inv act},
\beq
w_x=aG_x(\br) \, \sep w_y=aG_y(\br)  \,,
\label{-field}
\eeq

Using the dislocation equation of motion  \eqref{vdis}, together with our expressions \eqref{+field} and  \eqref{-field} for the fields at the $+$ and $-$ dislocations, , we obtain the equations of motion for the dislocations:
\beqn
{dx_\pm\over dt}=\mp\mu_xaG_y(\br) +f^\pm_x(t) \,,
\nn\\ 
{dy_\pm\over dt}=\pm\mu_yaG_x(\br) +f^\pm_y(t) \,,
\label{pairEOM}
\eeqn
where, as we did for the rotation invariant case, we have added random noises $\bff^\pm$ to the equation of motion for each dislocation, to take into account random microscopic processes (including, but not limited to, thermal fluctuations) that move the dislocations. We will continue to take these noises to have correlations given by \ref{disforcecorr}, with the indices $\alpha$ and $\beta$ running over the two values $+$ and $-$.

From \ref{pairEOM},  it follows that the relative displacement $\br$ obeys 
\beqn
{dx\over dt}&&={dx_+\over dt}-{dx_-\over dt}=-2\mu_xaG_y(\br)+f_x(t) \,,
\nn\\ 
{dy\over dt}&&={dy_+\over dt}-{dy_-\over dt}=2\mu_yaG_x(\br)+f_y(t) \,,
\label{relEOM}
\eeqn
where the ``relative force" $\bff=\bff^+-\bff^-$. From this, it follows that the relative force is also Gaussian, with mean and variance given by:
 \beqn
 \la\bff(t)\ra&&={\bf 0} \,\nn,\\
 \la f_i(t) f_j(t') \ra&&=2\Bigg(\Delta_x \delta_{ij}^x+\Delta_y \delta_{ij}^y\Bigg)\delta(t-t')\delta_{\alpha\beta}\,.\nn\\
\label{reldisforcecorr}
 \eeqn

Using our earlier expression \eqref{rot inv fields}  relating the Greens functions $G_{x,y}$ to the gradient of the  potential \eqref{U active}, we can rewrite this as
\beqn
{dx\over dt}=-2{\mu_x\over\varpi}a\pp_xV(\br)+f_x(t)  \,,
\nn\\ 
{dy\over dt}=-2\mu_y\varpi a\pp_yV(\br)+f_y(t)  \,.
\label{relEOMpot}
\eeqn

Note that {\it if} the noise variances $\Delta_{x,y}$ and the effective ``relative mobilities" 
\beq
\mu^{\rm rel}_x\equiv {2a\mu_x\over\varpi}\sep \mu^{\rm rel}_y\equiv 2a{\mu_y\varpi}
\label{mudef}
\eeq
satisfied
\beq
{\Delta_x\over\Delta_y}={\mu^{\rm rel}_x\over\mu^{\rm rel}_y}  \,,
\label{fdt}
\eeq
then the equations of motion \ref{relEOMpot} could be written in the form
\beqn
{dr_i\over dt}=-\tilde\Gamma_{ij}\pp_jU(\br)+f_i(t)  \,, 
\label{relEOMeq}
\eeqn
with $U(\br)=KV(\br)$, 
a diagonal kinetic coefficient tensor
 \beqn
\tilde\Gamma_{ij}=\Bigg(\tilde\Gamma_x \delta_{ij}^x+\tilde\Gamma_y \delta_{ij}^y\Bigg)\,,
\label{kinten}
 \eeqn
and the noise correlations given by 
 \beqn
 \la f^\alpha_i(t) f^\beta_j(t') \ra&&=2\kbt\tilde\Gamma_{ij}\delta(t-t')\delta_{\alpha\beta}\,.\nn\\
\label{disforcecorreqeq}
 \eeqn
 The reader can easily show for herself that this works provided the temperature $T$, kinetic coefficient tensor components $\tilde\Gamma_{x,y}$, and effective spin wave stiffness $K$ obey
 \beq
 {\kbt\over K}={\Delta_x\over\mu^{\rm rel}_x}={\Delta_y\over\mu^{\rm rel}_y} \sep \tilde\Gamma_{x,y}={\mu^{\rm rel}_{x,y}\over K} \,.
 \eeq
 
 The relation \ref{disforcecorreqeq} between these noise correlations and the kinetic coefficient tensor 
 $\tilde\Gamma_{ij}$ is exactly that required by the fluctuation-dissipation relations\cite{chaikin}. Therefore, the relative motion of the dislocation pair is exactly that of a pair moving in {\it equilibrium} in the potential $U(\br)$. This is precisely the equilibrium model for the Kosterlitz-Thouless transition discussed in section [\ref{disloeqsymbreak}]. Therefore, {\it if} this system {\it did} happen to satisfy the relation \eqref{fdt}, it would undergo a Kosterlitz-Thouless unbinding transition as noise was increased. Most importantly, at sufficiently small, but non-zero, noise (i.e., sufficiently low temperature in the equilibrium model), dislocations 
 would remain bound, and the active smectic phase 
 would be stable.
 
 However, in a {\it weak} symmetry breaking field, the system will always be far from satisfying the relation \eqref{fdt}. This can be seen from limiting behaviors of the $\Delta$'s and $\mu$"s in the limit of weak symmetry breaking field $g$: they all go to finite constants as $g\to0$ {\it except} $\mu_y$, which, as discussed earlier, vanishes according to $\mu_y\propto g$ as $g\to 0$.
 
 Even if the symmetry breaking field is {\it not} weak, there is no reason our non-equilibrium active smectic should obey \eqref{fdt}. Therefore, the connection to the equilibrium Kosterlitz-Thouless transition just made for the special case in which equation \eqref{fdt} is satisfied will not, in general, hold.
 
 Nonetheless, we expect this system to undergo something very like a Kosterlitz-Thouless transition, at least for very weak symmetry breaking fields. To establish this, however, we cannot use equilibrium arguments. Instead, we must use the more general Fokker-Planck equation for the generic non-equilibrium system, to which we now turn.

Using standard techniques\cite{chaikin}, we can show that the stochastic equation of motion \eqref {relEOMpot} for the relative displacement $\br$ of the two dislocations
implies that the probability density $\rho(x,y,t)$ for the relative displacement vector $\br$ obeys the Fokker-Planck equation:
\beq
\pp_t\rho+\vnab\cdot(\bu\rho)-(\Delta_x\pp_x^2+\Delta_y\pp_y^2)\rho=0\,,
\label{FP0}
\eeq
where 
\beq
u_i(\br)\equiv-\mu_{ij}\pp_jV(\br) \,,
\label{udef}
\eeq
with the effective relative mobility tensor 
\beq
\mu_{ij}=\Bigg(\mu^{\rm rel}_x \delta_{ij}^x+\mu^{\rm rel}_y \delta_{ij}^y\Bigg)\,,
\label{muten}
\eeq
is the deterministic part of the relative dislocation velocity in equation \eqref{relEOMpot}.

Looking for steady-state solutions to this equation, we can set the time derivative to zero, and drop the time dependence of $\rho(x,y,t)$ (that is, set $\rho(x,y,t)=\rho(x,y)$). Doing so, and using our expression \eqref{U active} for $V(\br)$,   leads to the steady state equation:
\bew
\beq
(\Delta_x\pp_x^2+\Delta_y\pp_y^2)\rho+\pp_x\left({\gamma_xx\rho\over x^2+\alpha y^2}\right) +\pp_y\left({\gamma_yy\rho\over x^2+\alpha y^2}\right)=0\,,
\label{FP1}
\eeq
\ew
where we've defined 
\beq
\alpha\equiv{D_x\over D_y}={1\over\varpi^2}
\label{alphadef}
\eeq
 and 
\beq
\gamma_i={\mu^{\rm rel}_i\over2\pi} \sep\,\, i=(x,y) \,.
\label{gammadef}
\eeq

Since we expect that, for small symmetry breaking field $g$, $\mu_y\propto g$, while all other parameters should 
go to finite, non-zero constants
as $g\to0$, we therefore expect 
\beq
\gamma_y\propto g \,\,\, {\rm as} \,\,\,\,  g\to0\,.
\label{gammascale}
\eeq
We will use this scaling law later to determine how the critical noise strength $\Delta_y^c(\zeta, g)$ at which noise destroys smectic order depends on activity $\zeta$ and symmetry breaking field strength $g$.

Our experience with the equilibrium case suggests that we seek a solution of this equation which falls off like a power law;  i.e.,  
$
\rho(\br)\propto r^{-\nu}$ for large $r$. We will therefore insert the scaling  ansatz
\beq
\rho(x,y)=y^{-\nu} \Phi(x/y) \,
\label{rhoscale}
\eeq
into \ref{FP1}.

Doing so, we find that such a solution will indeed work, provided that the scaling function $\Phi(z)$ obeys the ODE:
\bew
\beq
\Delta_x\Phi''(z)+\Delta_y\bigg[z^2\Phi''(z)+2(\nu+1)z\Phi'(z)+\nu(\nu+1)\Phi(z)\bigg]=-\gamma_x{d\over dz}\left({z\Phi(z)\over z^2+\alpha}\right) +\gamma_y\left[{(1+\nu)\Phi(z)\over z^2+\alpha}+z{d\over dz}\left({\Phi(z)\over z^2+\alpha}\right)\right]\,.
\label{scalingODE}
\eeq
\ew

We will not need to solve this equation (fortunately!); rather, we will use it simply to establish that the scaling function $\Phi(z)$ does nothing singular as $\gamma_y\to0$; that is, as the symmetry breaking field strength goes to zero. In particular, the {\it range} of $\Phi(z)$ - that is, the value of $z$ at which $\Phi(z)$ starts falling off fast enough that its integral converges - does not diverge as $\gamma_y\to0$. To see this, consider \eqref{scalingODE} at $z=0$, where it reads
\bew
\beq
\Phi''(z=0)=-\bigg[\left({\Delta_y\over\Delta_x}\right)\nu(1+\nu)+{(\gamma_x-\gamma_y(1+\nu))\over\alpha\Delta_x}\bigg]\Phi(z=0)\,.
\label{scalingODE}
\eeq
\ew

Since $\Delta_x$ and $\Delta_y$ are of the same order of magnitude, the coefficient of $\Phi(z=0)$ in \eqref{scalingODE} is at least $O(1)$. Therefore,  $\Phi''(z=0)$ is a negative number whose magnitude is $\gtrsim O(1)$ times $\Phi(z=0)$. This means that we would expect $\Phi(z)$ to drop from $\Phi(z=0)$ to small values on a scale no larger than $O(1)$. For large $z$ the behavior of $\Phi(z)$  obeys  $z^2\Phi''(z)+2(\nu+1)z\Phi'(z)+\nu(\nu+1)\Phi(z)=0$, which implies $\Phi(z)\sim z^{-\nu}$ for large $z$.

Our scaling ansatz \eqref{rhoscale} implies that
\beq
f(y)\equiv\int_{-\infty}^\infty \rho(x,y) dx=\int_{-\infty}^\infty y^{-\nu} \Phi(x/y) dx=y^{1-\nu}\Upsilon_1 \,,\label{fdef}
\eeq
where we've defined 
\beq
\Upsilon_1\equiv\int_{-\infty}^\infty \Phi(z) dz \,.
\label{wdef1}
\eeq

Our above observation that  $\Phi(z)$ becomes small for $z\gtrsim \cO(1)$ regardless of the value of $\gamma_y$ implies that $\Upsilon_1=\cO(1)$ regardless of the value of $\gamma_y$ as well. Note also that $\Upsilon_1$ is independent of $y$.
 
Now, integrating 
equation \eqref{FP1} over $x$ from $-\infty$ to $\infty$ gives
\beq
\Delta_yf''(y)=-\gamma_y{d\over dy}\bigg[y \int_{-\infty}^\infty {\rho(x,y) dx\over x^2+\alpha y^2}\bigg]\,.
\label{FP2}
\eeq
In deriving this equation, we have dropped surface terms that must vanish since the scaling function $\Phi(z)$ vanishes sufficiently rapidly to be integrable.

Using our scaling ansatz \eqref{rhoscale} enables us to rewrite 
\beq
\int_{-\infty}^\infty {\rho(x,y) dx\over x^2+\alpha y^2}=\int_{-\infty}^\infty {y^{-\nu} \Phi(x/y) dx\over x^2+\alpha y^2}=y^{-1-\nu} \Upsilon_2
\label{int2}
\eeq
where we've defined
\beq
\Upsilon_2\equiv\int_{-\infty}^\infty {\Phi(z) dz\over z^2+\alpha} \,.
\label{w2def}
\eeq
Like $\Upsilon_1$, $\Upsilon_2$ can be readily seen to be $\cO(1)$, even when $\gamma_y\to0$.
Thus \ref{FP2} can be rewritten as
\beq
\Delta_y{d^2\over dy^2}(\Upsilon_1y^{1-\nu})=-\gamma_y{d\over dy}(\Upsilon_2y^{-\nu})\,,
\label{FP3}
\eeq
which can readily be seen to work (because both sides scale the same way with $y$; specifically, like $y^{-(\nu+1)}$) provided
\beq
\Delta_y\Upsilon_1\nu(\nu-1)=\gamma_y\Upsilon_2\nu\,.
\label{FP4}
\eeq
Since $\nu=4$ at the dislocation unbinding transition, \eqref{FP4} implies that the value  $\Delta_y^c$ of $\Delta_y$ at the transition obeys
\beq
\Delta_y^c=\gamma_y\left({\Upsilon_2\over3\Upsilon_1}\right)\,.
\label{Deltayc}
\eeq
Since $\Upsilon_1$ and $\Upsilon_2$ are of $\cO(1)$, even when $\gamma_y\to0$, \eqref{Deltayc} implies that
\beq
\Delta_y^c=\gamma_y\times\cO(1)\,.
\label{Deltayc2}
\eeq

Given our earlier argument that $\gamma_y$ should vanish linearly with the symmetry breaking field $g$ as $g\to0$, this result implies that the 
noise strength $\Delta_y^c$ at the transition should also vanish linearly with $g$ as $g\to0$. 
 
Since we also expect that the noise strengths $\Delta_{x,y}$ in the dislocation equation of motion noise correlations \eqref{reldisforcecorr} are both proportional to the spin wave noise strength  $\Delta$, this implies that the value $\Delta_c$ of the spin wave noise strength $\Delta$ at the transition should also scale linearly with the symmetry breaking field strength $g$ for small $g$; that is
\beq
\Delta_c\propto g \,\,\, {\rm as} \,\,\,\,  g\to0\,.
\label{noisescale}
\eeq
This result should be contrasted with the equilibrium result 
\beq
\Delta_c^{\rm eq}\propto \sqrt{g} \,\,\, {\rm as} \,\,\,\,  g\to0\,,
\label{noisescale eq}
\eeq
which follows from equation \eqref{TKTscale} if we interpret $T_{KT}$ as the critical noise correlation strength. We therefore see that, for small symmetry breaking fields $g$, the critical noise strength $\Delta_c$ for dislocation unbinding and the melting of the smectic phase is much weaker for active smectics than for equilibrium smectics. Active smectics are {\it less} robust against melting, even in the presence of symmetry breaking fields, than their equilibrium counterparts.

Another consequence of this result \ref{noisescale} is that the critical value $\eta_c$ of the exponent $\eta$ for algebraic decay of smectic translational order becomes non-universal. Indeed, since the diffusion constants $D_x$ and $D_y$ go to finite, non-zero constants as the symmetry breaking field $g\to0$, while the critical noise strength $\Delta_c$ vanishes according to \ref{noisescale}, the value of $\eta_c$ also vanishes linearly with symmetry breaking field:

\beq
\eta_c\propto {\gamma_y\over\sqrt{D_xD_y}}\propto g \,\,\, {\rm as} \,\,\,\,  g\to0\,.
\label{etascale}
\eeq

\section{From Malthusian to incompressible active smectics}\label{from}

In the preceding paragraphs, for the sake of simplicity we have ignored the existence of a local velocity field 
$\bv(\br)$. This field has two effects:  it can advect the dislocations,  and  it can  modify the expressions of $w_x(\br)$ and $w_y(\br)$. The equations of motion for the dislocation pair separation now read
\beqn
{dx\over dt}&&=2   \psi v_x(\br)-2\mu_xw_y(\br)+f_x(t) \,,
\nn\\ 
{dy\over dt}&&=2   \psi v_y(\br)+2\mu_yw_x(\br)+f_y(t) \,,
\label{relEOMV} 
\eeqn
  where by $\bv(\br)$ we mean the velocity field at the point $\br$ generated by a $+1$ dislocation, and the factor of $2$ comes from the fact that an equal and opposite velocity field is generated by the $-1$ dislocation at the position of the $+1$ dislocation, and the two effects on the relative motion add. 
  Note that the prefactors of $v_x(\br)$ and $v_y(\br)$  are not $2$, as one might naively expect, but $2\psi$, where the factor 
$\psi$ captures the effects of drag between the dislocations and the substrate, as discussed in the appendix.
However, these effects do not modify our analysis since we will show that the velocity contribution is subdominant compared to that of $w_x, w_y$. 
The equations for the velocity field involve the force balance, the layer dynamics equation and the density {\it non conservation } equation. At steady state the layer dynamical equation \eqref{layerdyneq} becomes, see Appendix \ref{appendix}:
\begin{eqnarray}
0 = v_y(\br)+D_y \partial_y w_y + D_x \pp_x w_x + f, \label{vysteadystate}
\label{sweomsymbreakV}
\end{eqnarray} 
This equation shows that $\bv(\br)$ is of the order of derivatives of $\partial_x w_x, \partial_y w_y$ which lead to
terms subdominant compared to $w_x, w_y$ in the long distance limit, hence our claim that the advection term in \eqref{relEOMV} is subdominant.
The force balance expresses the fact that the momentum extracted from the substrate is balanced 
by the divergence of the stress. The momentum exchange with the substrate involves not only the usual friction term, but because we are dealing with an active system, also involves gradients of layer spacing plus bend and splay of the layer normal. The stress tensor involves  the layer compression term as in conventional smectics, the pressure term and the usual active stress of anisotropic systems. The viscous term which is of higher order in gradients than 
substrate friction may be omitted.
The density balance equation involves a source term which expresses that whenever the pressure departs from its homeostatic value, the elements building the active smectic are either created or destroyed. It has been shown that under these conditions on long time scales one can replace the pressure term by a bulk viscous term \cite{homeoTissuetwo}. As already pointed out viscous terms can be omitted in the long wavelength limit and it is then straightforward to show that the equation we found for  $w_x, w_y$ given in Eq. (\ref{steady state}) 
 is valid and the effect of flow is simply to renormalize the coefficients. Thus our conclusions concerning the Malthusian case are valid even when one takes into account a momentum exchange with the substrate 
 more complex than just friction.

We now turn to the incompressible active smectic case. The pressure becomes a Lagrange multiplier which can be calculated with the condition that the divergence of the velocity field vanishes. Other equations are identical to the one we just discussed. They can be solved in a straightforward way. All together, this adds up to 
eight parameters in the problem. Following the same logic as before we obtain:
\beqn
&&w_x(\br)=-{y\gamma^{(1)}_x \over 2\pi(\alpha_1 y^2+x^2)}+{y\gamma^{(2)}_x \over 2\pi(\alpha_2 y^2+x^2)}\,,\\
&&w_y(\br)={x\gamma^{(1)}_y \over  2\pi(\alpha_1 y^2+D_yx^2)}-{x\gamma^{(2)}_y \over  2\pi(\alpha_2 y^2+D_yx^2)}\,. 
\nonumber\\
\label{dislocsymbreakactV}
\eeqn

We have obtained expressions  for $\gamma^{(1)}_x, \gamma^{(2)}_x, \gamma^{(1)}_y, \gamma^{(2)}_y, \alpha_1, \alpha_2$  in terms of the aforementioned  8 parameters,  but they are not very illuminating and so we will not give them here. In the large friction regime one recovers the results of the Malthusian case, with $\gamma^{(2)}_x, \gamma^{(2)}_y$ going to zero, $\gamma^{(1)}_x, \gamma^{(1)}_y$ going to $\gamma_x, \gamma_y$ and $\alpha_1, \alpha_2$ going to $\alpha$. The important point is that the expressions \ref{dislocsymbreakactV} appear as the difference between two functions having the same structure as the one appearing in the Malthusian case, with the same scaling. Thus there is an important parameter space for which the sign of $w_x(\br), w_y(\br)$ and the scaling are the same as in the Malthusian case. This means that the same procedure as in the preceding section can be followed and that the conclusions detailed in the Malthusian case are valid in general in a broad range of parameters even in the incompressible case.

\section{Summary, conclusions, and suggestions for future work}\label{sum}

We have shown that dislocations in dry active Malthusian smectics - that is, smectics lacking all conservation laws, including that of particle number-  behave very differently from those in equilibrium smectics. Specifically:

\noindent 1) they can move spontaneously, even in isolation.

\noindent 2) Because of this, active smectics with ``constant stress" boundary conditions can never reach a steady state. Instead, they either grow forever, or shrink and disappear. This behavior is similar to that of tissues\cite{tissue}.

\noindent 3) When their boundaries are fixed, active smectics reach a state of ``homeostasis", in which the spontaneous motion of isolated dislocations ceases.

\noindent 4) However, even in the state of homeostasis, dislocations are always unbound in rotation invariant active smectics, if there is any noise, however small. This means that the active smectic phase does not, in fact, exist at finite noise.

\noindent 5) By applying rotational symmetry breaking fields, active smectics can be stabilized against dislocation unbinding for sufficiently small noise. However, for weak symmetry breaking fields, active smectics are less robust against noise than their equilibrium counterparts. Specifically, the critical noise strength $\Delta_c$ above which dislocations unbind and smectic order is lost scales linearly with the symmetry breaking field strength $g$, in contrast to the $\sqrt{g}$ scaling of the critical temperature in an equilibrium smectic.

\noindent 6) As a result, the exponent $\eta$ for the algebraic decay of smectic correlations (given by equation \eqref{etaxyactive}) becomes non-universal at melting, and in fact vanishes linearly with symmetry breaking field strength $g$ as $g\to0$. This should be contrasted with the universal value $\eta(T_c)=1/4$ of this exponent in equilibrium smectics with a symmetry breaking field, a result that is completely independent of the symmetry breaking field $g$.

Our work here has, of course, focused mainly on a very particular type of smectic, namely dry Malthusian apolar smectics. In contrast to equilibrium systems\cite{chaikin}, for which the exact nature of the dynamics - in particular, what conservation laws the dynamics respects - has no effect on the equal time correlations of smectic fluctuations, in active systems, because they are non-equilibrium, all we have is dynamics, and, so, {\it a priori}, results could change if we consider smectics with conservation laws. The most obvious of these is number conservation, which  our Malthusian smectics lack, due to ``birth and death" of the constituent active particles. One can of course imagine many situations in which birth and death are absent (at last on the time scale of an experiment). Such systems will have very different hydrodynamic equations, because the conserved particle density will now become a slow, hydrodynamic variable\cite{MPP}, thereby completely changing the dynamics. The spin wave theory for this case has already been worked out \cite{apolarsm}, 
 and our analysis of the incompressible case suggests our results are quite general.

Likewise, momentum conservation, which will hold for freely suspended 2d systems (i.e.,
those not in contact with a substrate to which they can lose momentum)
always radically changes the dynamics, for similar reasons.

Finally, polarity (ie., the absence of head-tail symmetry) is also known\cite{polarsm}
to change the nature of the spin-wave theory of active smectics. So its role in dislocation behavior should also be investigated.

We strongly suspect that in all of these more complicated systems, our fundamental conclusion - namely, that dislocations in a rotation invariant 2d active smectic will always be unbound in the presence of noise, meaning that the active smectic phase cannot exist at finite noise in two dimensions. We suspect this because, whatever the conserved quantities, rotation invariance forbids motion of dislocations, other than Brownian motion driven by noise  , or motion driven by {\it curvature} of the layers, in the direction perpendicular to the layers in a rotation invariant system.   Because the curvature field induced by dislocations will always fall off quite rapidly with distance from the dislocation, the motion induced by them will always be insufficient to bind a dislocation pair. Instead, dislocations can always unbind by diffusing apart in the direction normal to the layers, as we found here for dry Malthusian active smectics.

But, obviously, this conclusion is at best speculative until those other systems enumerated above are explicitly investigated.

\begin{acknowledgments}
JT thanks  The Higgs Centre for Theoretical Physics at the University of Edinburgh for their hospitality and support while this work was in progress. He likewise thanks the Max Planck Institute for the Physics of Complex Systems, Dresden, Germany, for their support through the Martin Gutzwiller Fellowship. JP thanks A. Bershadsky for drawing his attention on smectic order in fibroblast cells.

\end{acknowledgments}

\appendix

\section{Theory of active smectics} \label{appendix}

We describe the smectic configuration by the vector field $m \hat {\bf n}$ which also defines
 ${\bf w}(\br,t)=(a_0 m-1) \hat {\bf n}$, where $m=1/a$ is the density of layers, $\hat {\bf n}$
is a unit vector normal to the layers and $a_0$ a reference layer spacing. 
The vector field $m \hat {\bf n}$ satisfies a
general balance equation
\begin{equation}
\partial_t (m \hat {\bf n})+\nabla J_m = -\hat z \times {\bf J}_d \quad . \label{balanceeq}
\end{equation}
Here, the dislocation current is
\beq
{\bf J}_d=\sum_\alpha b_\alpha {\bf v}_\alpha \delta(\br -\br_\alpha) \quad ,\label{dislocationcurrent}
\eeq
where ${\bf v}_\alpha$ is the velocity of disloctation $\alpha$ and $b_\alpha$ its Burgers number.
The layer rate $J_m$ can in the absence of dislocations be identified with
the rate of change of layer displacement:
\beq
 J_m a_0=- \partial_t u    \sep {\rm no} \,{\rm dislocations}\,.
 \eeq
 Eqns. (\ref{balanceeq}) and (\ref{dislocationcurrent}) imply Eq. (\ref{burgers dif}).
 
 In a hydrodynamic theory, we write phenomenological constitutive equations
 for $J_m$ and for the force balance, using terms allowed by symmetry
 at lowest order in spatial derivatives. For an 
 up-down and rotationally symmetric smectic, these can be written as
 \beqn
 J_m&=&m {\bf v}\cdot \hat {\bf n}-D_m \hat {\bf n}\cdot \nabla m+\lambda_m \nabla \cdot \hat {\bf n} \label{Jm} \\
 \nabla \cdot \boldsymbol{\sigma} &=& \boldsymbol{\mu}\cdot {\bf v}+ \boldsymbol{\nu}\cdot \nabla m+\lambda_b^v (\hat {\bf n}\cdot  \nabla) \hat {\bf n} \nonumber \\
& +&\lambda_s^v \hat {\bf n} (\nabla \cdot \hat {\bf n}) \label{veq}
 \eeqn
where $\bf v$ is the material flow velocity and we have introduced phenomenological coefficients $D_m$, $\lambda_m$, $\lambda_{b,s}^v$ as well
as mobility and kinetic tensors $\boldsymbol{\mu}$, $\boldsymbol{\nu}$ and we have omitted the noise for simplicity.

The stress tensor $\boldsymbol{\sigma}$  also obeys a constitutive relation which is of the form
\beq
\sigma_{ij}= -P \delta_{ij} -B \bigg(\hat n_i \hat n_j-\frac{1}{2}\delta_{ij}\bigg)\frac{m-m_0}{m_h} \quad ,\label{sigma}
\eeq
where $P$ is pressure and we have omitted higher order and viscous terms which are subdominant in the hydrodynamic limit.
To study small deformations, we write $m=1/a_0+\delta m$ and $\hat {\bf n}=\sin(\phi) \hat x-\cos(\phi)\hat y$.
To linear order in $\phi$ and $\delta m$ we then have 
\beq
a_0 m \hat {\bf n}\simeq   \phi \hat x-(1+a_0\delta m)\hat y \quad ,
\eeq
which is the same as Eq. (\ref{wdef}). Furthermore, we have to lowest order $w_x\simeq \phi$ and 
$w_y \simeq -a_0 \delta m$. Finally, Eq. (\ref{Jm}) becomes
\beq
-J_m a_0 = v_y+D_x \partial_x w_x + D_y \partial_y w_y \label{Jmlin}
\eeq
where $D_y=D_m$ and $D_x=-\lambda_m a_0$. 
In steady state $J_m=0$ and
we obtain 
Eq. (\ref{vysteadystate}). 

The velocity $v_y$ can be determined
using Eqns. (\ref{veq}) and (\ref{sigma}). 
Doing so, we have to distinguish
the Malthusian from the incompressible case. In the Malthusian case, material is not conserved and
$\nabla \cdot {\bf v}$ does not vanish. In this case, $P=-\eta_b  \nabla \cdot {\bf v}$, where
$\eta_b$ is a bulk viscosity and the contributions from viscosity and pressure are irrelevant
in the hydrodynamic limit.
We then have
\beq
\mu_{yy} v_y \simeq  \left(\lambda_s^v-B\left(\frac{m-m_0}{m_h}\right)\right)\partial_x \phi-\left(\nu_{yy} + \frac{B}{2m_h}\right) \partial_y m 
\quad .
\eeq
Using this expression in (\ref{Jmlin}) to eliminate $v_y$
 we arrive in steady state at Eq. (\ref{steady state}) of a Malthusian active smectic
but with renormalized coefficients 
\beqn
D_x&=&-\lambda_m a_0+\frac{\lambda_s^v-B(m_h-m_0)/m_h}{\mu_{yy}} \\
D_y&=&D_m+\frac{\nu_{yy}+B/(2m_h)}{a_0\mu_{yy}} \quad .
\eeqn
In the incompressible case the
pressure acts as a Lagrange multiplier to impose the incompressibility constraint $\nabla\cdot {\bf v}=0$. This modifies the
hydrodynamic modes, see section \ref{from}.

The velocity of dislocation $\alpha$ can also be written on symmetry grounds to lowest order as
\begin{equation}
{\bf v}_\alpha=\psi{\bf v}+B'\left(\frac{m-m_h}{m_h}\right)b_\alpha \hat {\bf n}\times \hat z
\end{equation}
where $m_h$ is the layer density in the homeostatic state and which is equivalent to Eq. (\ref{selfp1}).

In the presence of a symmetry breaking field ${\bf H}$ rotation invariance is broken. In this
case additional symmetry breaking terms involving a tensor $S_{ij}=g \hat H_i \hat H_j$ are permitted.
These additional terms do not affect Eq. (\ref{Jmlin}) at linear order, see Eq. (\ref{vysteadystate}).
However, $S_{ij}$ allows for an additional term in the dislocation velocity
when rotation invariance is broken
\begin{equation}
{\bf v}_\alpha=  \psi{\bf v}+B'\left(\frac{m-m_h}{m_h}\right)b_\alpha \hat {\bf n}\times \hat z+ B''b_\alpha {\bf S}\cdot \hat {\bf n}\times \hat z
\end{equation}
where the field $\bf H$ is aligned such that ${\bf S}=gH^2 \hat x\otimes \hat x$, 
which corresponds to Eq. (\ref{selfp2}).
The dislocation velocities thus obey Eq. (\ref{vdis}) with $\mu_x =B'b_\alpha$ and $\mu_y=gH^2 B''b_\alpha$
and we obtain Eq. (\ref{muyscale}).

\end{document}